\documentclass[acmsmall,nonacm,screen]{acmart}
\usepackage{graphicx} %
\usepackage{xcolor}         %
\usepackage{subcaption}
\usepackage{longtable}
\newcommand{\cn}[1]{\textbf{\textcolor{red}{CITE}}}
\newcommand{\todo}[1]{\textbf{\textcolor{blue}{TODO}}}
\setcopyright{none}
\usepackage{xcolor}
\usepackage{tcolorbox}
\usepackage{listings}
\usepackage{caption}
\usepackage{tikz}
\usetikzlibrary{positioning, arrows.meta, fit, backgrounds}

\definecolor{dkviolet}{rgb}{0.3,0,0.5}
\definecolor{dkgreen}{rgb}{0,0.5,0}
\definecolor{ltblue}{rgb}{0,0.1,0.7}
\definecolor{dkblue}{rgb}{0,0.1,0.7}
\definecolor{dkred}{rgb}{0.7,0,0}

\lstdefinelanguage{Coq}{
    mathescape=true,
    texcl=false,
    morekeywords=[1]{Section, Module, End, Require, Import, Export,
        Variable, Variables, Parameter, Parameters, Axiom, Hypothesis,
        Hypotheses, Notation, Local, Tactic, Reserved, Scope, Open, Close,
        Bind, Delimit, Definition, Let, Ltac, Fixpoint, CoFixpoint, Add,
        Morphism, Relation, Implicit, Arguments, Unset, Contextual,
        Strict, Prenex, Implicits, Inductive, CoInductive, Record,
        Structure, Canonical, Coercion, Context, Class, Global, Instance,
        Program, Infix, Theorem, Lemma, Corollary, Proposition, Fact,
        Remark, Example, Proof, Goal, Save, Qed, Defined, Hint, Resolve,
        Rewrite, View, Search, Show, Print, Printing, All, Eval, Check,
        Projections, inside, outside, Def},
    morekeywords=[2]{forall, exists, exists2, fun, fix, cofix, struct,
        match, with, end, as, in, return, let, if, is, then, else, for, of,
        nosimpl, when},
    morekeywords=[3]{Type, Prop, Set, true, false, option},
    morekeywords=[4]{pose, set, move, case, elim, apply, clear, hnf,
        intro, intros, generalize, rename, pattern, after, destruct,
        induction, using, refine, inversion, injection, rewrite, congr,
        unlock, compute, ring, field, fourier, replace, fold, unfold,
        change, cutrewrite, simpl, have, suff, wlog, suffices, without,
        loss, nat_norm, assert, cut, trivial, revert, bool_congr, nat_congr,
        symmetry, transitivity, auto, split, left, right, autorewrite},
    morekeywords=[5]{by, done, exact, reflexivity, tauto, romega, omega,
        assumption, solve, contradiction, discriminate},
    morekeywords=[6]{do, last, first, try, idtac, repeat},
    morecomment=[s]{(*}{*)},
    showstringspaces=false,
    morestring=[b]",
    tabsize=3,
    extendedchars=false,
    sensitive=true,
    breaklines=false,
    basicstyle=\small,
    captionpos=b,
    columns=[l]flexible,
    identifierstyle={\ttfamily\color{black}},
    keywordstyle=[1]{\ttfamily\color{dkviolet}},
    keywordstyle=[2]{\ttfamily\color{dkgreen}},
    keywordstyle=[3]{\ttfamily\color{ltblue}},
    keywordstyle=[4]{\ttfamily\color{dkblue}},
    keywordstyle=[5]{\ttfamily\color{dkred}},
    stringstyle=\ttfamily,
    commentstyle={\ttfamily\color{dkgreen}},
    literate=
    {\\emp}{$\emptyset$}1
    {\\.}{$\cdot$}1
    {\\forall}{{\color{dkgreen}{$\forall\;$}}}1
    {\\exists}{{$\exists\;$}}1
    {<-}{{$\leftarrow\;$}}1
    {->}{{$\rightarrow\;$}}1
    {<->}{{$\leftrightarrow\;$}}1
    {<==}{{$\leq\;$}}1
    {\#}{{$^\star$}}1
    {\\o}{{$\circ\;$}}1
    {\@}{{$\cdot$}}1
    {\/\\}{{$\wedge\;$}}1
    {\\\/}{{$\vee\;$}}1
    {~}{{$\sim$}}1
    {\@\@}{{$@$}}1
    {\\mapsto}{{$\mapsto\;$}}1
    {\\hline}{{\rule{\linewidth}{0.5pt}}}1
}[keywords,comments,strings]

\definecolor{generator}{RGB}{75, 161, 241}
\definecolor{ltgenerator}{RGB}{75, 161, 241}
\definecolor{dkgenerator}{RGB}{75, 161, 241}
\definecolor{mutator}{RGB}{174, 62, 201}
\definecolor{ltmutator}{RGB}{174, 62, 201}
\definecolor{dkmutator}{RGB}{174, 62, 201}
\definecolor{shrinker}{RGB}{241, 172, 75}
\definecolor{ltshrinker}{RGB}{241, 172, 75}
\definecolor{dkshrinker}{RGB}{241, 172, 75}
\definecolor{checker}{RGB}{224, 49, 49}
\definecolor{ltchecker}{RGB}{224, 49, 49}
\definecolor{dkchecker}{RGB}{224, 49, 49}
\definecolor{printer}{RGB}{159, 168, 178}
\definecolor{ltprinter}{RGB}{159, 168, 178}
\definecolor{dkprinter}{RGB}{159, 168, 178}

\newcommand{\overlayfull}[2]{\tcbox[on line,boxsep=0pt, boxrule=1pt,left=2pt,right=2pt,top=2pt,bottom=2pt, colframe=dk#1, colback=lt#1]{#2}}

\lstnewenvironment{rocqcodesmaller}{\lstset{language=Coq,mathescape=true,basicstyle=\ttfamily\footnotesize}}{}

\lstdefinelanguage{Rust}{
    morekeywords=[1]{fn, impl, for, struct, type, let, mut, pub, use, if, else, match, return, self, Self, as, where},
    morekeywords=[2]{bool, String, usize, f64, str, Vec, Option, Some, None, ControlFlow},
    morecomment=[l]{//},
    morestring=[b]",
    sensitive=true,
    showstringspaces=false,
    mathescape=true,
    columns=[l]flexible,
    keepspaces=true,
    basicstyle=\ttfamily\footnotesize,
    identifierstyle={\color{black}},
    keywordstyle=[1]{\color{dkviolet}},
    keywordstyle=[2]{\color{ltblue}},
    commentstyle={\color{dkgreen}},
    stringstyle={\color{dkred}},
}

\lstnewenvironment{rustcode}{\lstset{language=Rust}}{}

\tikzset{
  component/.style={
    minimum size=0.9cm,
    draw,
    line width=1.5pt,
    font=\Large\sffamily\bfseries,
    inner sep=0pt,
    fill=white
  },
  generator box/.style={component, draw=generator, text=generator},
  checker box/.style={component, draw=checker, text=checker},
  shrinker box/.style={component, draw=shrinker, text=shrinker},
  printer box/.style={component, draw=printer, text=printer, line width=1pt},
  mutator box/.style={component, draw=mutator, text=mutator},
  feedback box/.style={component, draw=feedback, text=feedback},
  combfeedback box/.style={component, draw=feedback, text=feedback, minimum size=0.7cm},
  size box/.style={component, draw=shrinker, text=shrinker, fill=shrinker!15, minimum size=0.8cm, font=\normalsize\bfseries},
  loop box/.style={
    draw=gray!70,
    dashed,
    line width=1pt,
    rounded corners=6pt,
    inner sep=8pt
  },
  loop region/.style={
    draw=gray!70,
    dashed,
    line width=1pt,
    rounded corners=6pt,
    fill=gray!8,
    inner sep=8pt
  },
  seed pool/.style={
    draw=seedpool,
    dashed,
    line width=1pt,
    rounded corners=4pt,
    inner sep=5pt
  },
  loop arrow/.style={
    ->,
    >={Stealth[length=2.5mm]},
    line width=0.7pt
  },
  arrow label/.style={
    font=\tiny\sffamily,
    inner sep=1pt
  },
  loop title/.style={
    font=\scriptsize\sffamily,
    text=gray!70
  }
}

\title{Ephyra: A Property-Based Testing Framework for the LLM Era}
\author{Alperen Keles}
\affiliation{%
  \institution{Datadog, Inc.}
  \city{New York}
  \country{USA}
}
\email{alp.keles@datadoghq.com}

\begin{document}

\begin{abstract}
Coding agents now write a growing share of software tests; however, property-based testing (PBT) libraries are designed with historical considerations.
We present Ephyra, a PBT library in Rust built on three concepts: customizability, observability and generator synthesis.
Ephyra lets users compose their own testing loops from generation, mutation and shrinking, and synthesize these components to satisfy functional predicates.
Applying Ephyra to 14 open source systems, we have found 349 bugs and reported 300 of them, in database management systems (SQLite, Turso, PostgreSQL, DuckDB, DataFusion), the verified policy engine (Cedar), the SQL equivalence prover (QED), the URL parser used in Node.js (ada-url), reimplementations of existing open source systems in Rust (coreutils, brush, gitoxide), and compilers (binaryen, OCaml, and OxCaml).
\end{abstract}

\maketitle

\section{Introduction} \label{sec:intro}

Large language models (LLMs) have significantly affected every phase of the software development life cycle (SDLC). Testing is no exception; the way software tests are read, written, and leveraged has been changing especially due to the growing popularity of spec-driven development~\cite{kiro} and the ability to run CEGIS-style~\cite{solarlezama2006sketching} program synthesis loops~\cite{sumner2026bunrust,carlini2026ccompiler,yang2026programbench} relying on comprehensive test suites. This abstract addresses two questions---predominantly the latter question---in the midst of these changes:

\begin{enumerate}
    \item Should, and how can, we leverage LLMs in property-based testing workflows?
    \item How should the design of PBT libraries change with respect to the changing software engineering practices?
\end{enumerate}

The first question is increasingly investigated in the literature and practice of PBT. The main interest is in using models to write property-based tests~\cite{vikram2024llmpbt,maaz2025agenticpbt,hegelzoo} with further research on LLM-guided generation~\cite{xia2024fuzz4all,deng2023titanfuzz} and LLM-authored test-case generators~\cite{zhang2025g2fuzz,yang2024whitefox,vikram2026gentoo,chen2025elfuzz} and test case shrinkers~\cite{zhang2024lpr}.
The second question, although naturally explored by using PBT libraries through agentic workflows, is currently under-researched. Similar questions have been posed in the context of dependency management~\cite{tomlinson2026libraries,willison2026porting,ronacher2025buildityourself} and data management systems domains~\cite{liu2026aioverlords}. In the context of PBT, the most significant effort to date is PBT skills~\cite{hegelskill,trailofbitspbtskill} for LLMs.
The core hypothesis behind the proposed library in \S\ref{sec:lib} is two-fold. First, we posit that the future of property-based testing is hybrid. We hypothesize that instead of simply sampling random inputs from a constructed distribution, the future PBT workloads will interleave generation, enumeration and mutation depending on the domain and system under test. With this hypothesis, we develop a programmable PBT~\cite{keles2026programmablepropertybasedtesting} library that exposes primitives for constructing testing loops.

Second, program synthesis for verifiable domains has shown promising results with the growing capabilities of LLMs~\cite{karpathy2026autoresearch,ouyang2025kernelbench,novikov2025alphaevolve}. Existing work on generator derivation strategies includes type-based generators~\cite{mista2018dragen,proptestderive}, specification-based generators~\cite{GeneratingGoodGenerators} derived from inductive specifications, and the randomness-parsing approach for deriving shrinkers and mutators from a given generator~\cite{maciver2020reducer,goldstein2023reflecting,maciver2026hegel}. However, these approaches are incompatible with arbitrary functional predicates, and randomness-parsing derives specific shrinkers/mutators that are immutable. Our library includes a synthesis tool that can optimize generators for a given set of metrics while adhering to the validity predicates as well as correct-by-construction shrinker/mutator synthesis that is validated against the generator.

Under these hypotheses, we propose three design principles for our new library, Ephyra\footnote{\url{https://github.com/datadog-labs/ephyra}}:

\begin{enumerate}
    \item Verifiable components should be synthesized.
    \item Customizability should be a first-class concern.
    \item Observability should be a first-class primitive.
\end{enumerate}

\section{Background}

This section presents existing literature and design space for the three principles outlined in \S\ref{sec:intro}.

\subsection{Components in PBT}

A \emph{component} is a PBT abstraction, such as a \textcolor{generator}{generator}, \textcolor{mutator}{mutator}, \textcolor{shrinker}{shrinker} or \textcolor{printer}{printer}, that defines an interface operating on the input type of the property. The classical property runner defined in QuickCheck~\cite{ClaessenH00} can be decomposed into the following property runner diagram:

\begin{figure}[h]
  \centering
  \resizebox{0.95\linewidth}{!}{\begin{tikzpicture}[node distance=0.8cm]
  \node[generator box] (G) {G};
  \node[checker box, right=1.5cm of G] (C1) {C};

  \draw[loop arrow] ([yshift=3pt]G.east) to[] node[arrow label, above] {\texttt{check e}} ([yshift=3pt]C1.west);
  \draw[loop arrow] ([yshift=-3pt]C1.west) to[] node[arrow label, below] {\texttt{gen}} ([yshift=-3pt]G.east);

  \begin{scope}[on background layer]
    \node[loop region, fit=(G)(C1), inner ysep=16pt, inner xsep=6pt, label={[loop title]above:Generation Loop}] (genloop) {};
  \end{scope}

  \node[shrinker box, right=1.5cm of C1] (S) {S};
  \node[checker box, right=1.5cm of S] (C2) {C};

  \draw[loop arrow] ([yshift=3pt]S.east) to[] node[arrow label, above] {\texttt{check e'}} ([yshift=3pt]C2.west);
  \draw[loop arrow] ([yshift=-3pt]C2.west) to[] node[arrow label, below] {\texttt{shrink e | e'}} ([yshift=-3pt]S.east);

  \begin{scope}[on background layer]
    \node[loop region, fit=(S)(C2), inner ysep=16pt, inner xsep=6pt, label={[loop title]above:Shrinking Loop}] (shrinkloop) {};
  \end{scope}

  \draw[loop arrow] (C1.east) -- node[arrow label, above] {\texttt{shrink e}} (S.west);

  \node[printer box, right=1.5cm of C2] (P) {P};

  \draw[loop arrow] (C2.east) -- node[arrow label, above] {\texttt{print e/e'}} (P.west);
\end{tikzpicture}}
  \caption{Abstract representation of QuickCheck runner taken from Programmable PBT~\cite{keles2026programmablepropertybasedtesting}.}
  \label{fig:quickchick-property-runner-loops}
\end{figure}
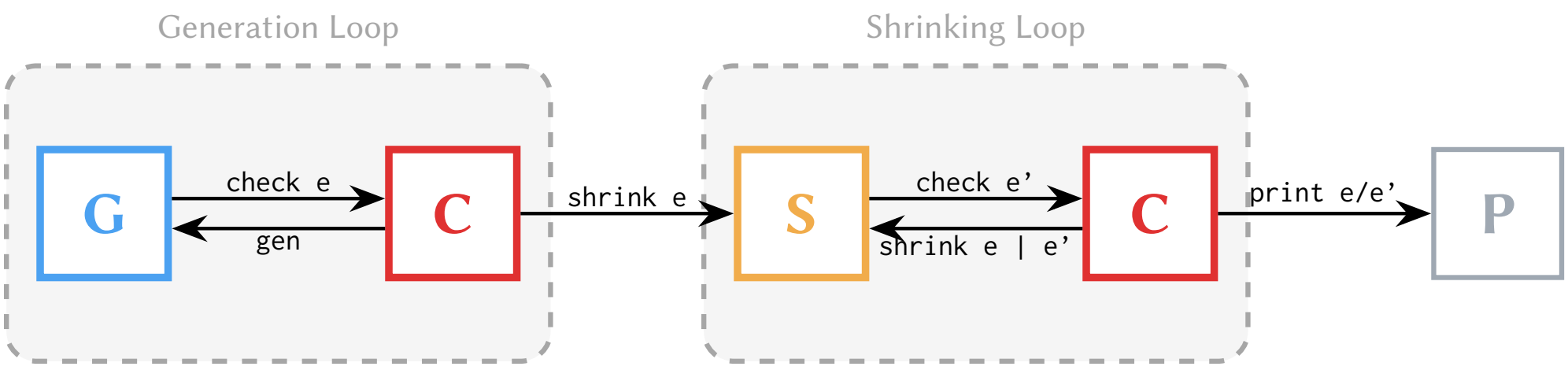

The implementation of these components is a significant part of the PBT development; the quality of the test case generator has a significant effect on bug finding performance~\cite{keles2026etnajfp}, and further so, the design of the libraries themselves is affected by this effort trade-off.
Many modern libraries use the ``randomness-parsing''~\cite{maciver2019hypothesis} approach to test case generation, in which they generate random bytes that are then interpreted into structures, citing the benefit of this approach in automatically producing correct-by-construction shrinkers and mutators~\cite{maciver2020reducer} without additional effort.
Other libraries provide ``derivations'', compile-time code generation via macros to produce generators and shrinkers based on a type~\cite{mista2018dragen,proptestderive}, or in the case of QuickChick, inductive specifications~\cite{paraskevopoulou2022computing}. Derivations are immensely useful in their convenience; derived generators can remove the hurdle of tediously writing and maintaining many small generators that could be trivial to write but time-consuming; derived shrinkers can be as good as the correct-by-construction shrinkers the randomness parsing libraries provide~\cite{keles2026evaluatingshrinking}.
Most of these derivations, however, are not long term testing strategies but starting points that lower the cost of property-based testing a new system. One notable exception is the derivation of generators that satisfy inductive specifications~\cite{GeneratingGoodGenerators}, giving rise to unlimited theoretical complexity, however limited in its applicability to testing programs due to the lack of such inductive specifications in practice.
In Ephyra, we close this gap by ``synthesizing'' generators that satisfy functional specifications like the validity of a SQLite file (\texttt{PRAGMA integrity\_check}) while optimizing the generator in a hill-climbing loop with respect to statistics we collect from the system under test.

\subsection{Customizability in PBT}

A property-based testing framework typically provides a language for writing properties, a set of APIs for writing generators, automations for deriving generators and shrinkers from the input types, one or more testing loops with limited configurability for testing the properties by sampling the provided generators, and some observability hooks for measuring statistics computed from the tests.
Programmable PBT~\cite{keles2026programmablepropertybasedtesting} is a new paradigm for PBT library design that gives the library users the primitives for constructing their own testing loops, a new axis in customization. The original work presents two design choices, (1) a property data structure in Rocq that carries a dependently-typed list of input values, and (2) a property data structure in Racket that carries a dynamically-typed context, both of which allow for inspecting the shape of the property.
Leveraging the capability of inspecting the property, the authors then build four APIs---building blocks---for constructing existing property runners in the literature such as below:

\begin{figure}[h!]
\centering
\begin{minipage}{0.65\linewidth}
\begin{rocqcodesmaller}
Definition runLoop (fuel : nat) (cprop : Prop \emp) :=
  let fix runLoop' (fuel : nat) (cprop : Prop \emp)
                   (passed : nat) (discards : nat) : G Result :=
    match fuel with
    | O => ret (mkResult discards false passed [])
    | S fuel' =>
      $\overlayfull{generator}{input <- gen cprop (log2 (passed + discards));;}$
      $\overlayfull{checker}{res <- run cprop input ;;}$
      match res with
      | Normal seed false => (* Fails *)
        $\overlayfull{shrinker}{let shrunk := shrinker 10 cprop seed in}$
        $\overlayfull{printer}{let printed := print cprop 0 shrunk in}$
        ret (mkResult discards true (passed + 1) printed)
      | Normal _ true => (* Passes *)
        runLoop' fuel' cprop (passed + 1) discards
      | Discard _ _ => (* Discard *)
        runLoop' fuel' cprop passed (discards + 1)
      end
    end in
  runLoop' fuel cprop 0 0.
\end{rocqcodesmaller}
\end{minipage}
\caption{QuickCheck-style property runner written in Rocq taken from Programmable PBT~\cite{keles2026programmablepropertybasedtesting}.}
\label{fig:ppbt-runloop}
\end{figure}

The programmability of Ephyra is limited compared to the original design; we inherit the four building blocks \texttt{generate}, \texttt{mutate}, \texttt{shrink}, and \texttt{run} but keep the property representation opaque. This hybrid design allows for the users to write their own property runners, but not the building blocks themselves. For instance, the users can arbitrarily interleave generation and mutation, but never at the granularity of individual input variables, only for the whole set of input variables to the property. We present the capabilities of the library in \S\ref{sec:lib}.

\subsection{Observability in PBT}

The first question one must ask once they have written a property-based test is: \emph{What does the test actually do?}
The typical mechanisms PBT libraries provide for understanding the generators are (1) sampling a number of inputs from the generators, and (2) computing aggregate statistics via \texttt{collect} callbacks.
Tyche~\cite{goldstein2024tyche} and the \texttt{OpenPBTStats}~\cite{goldstein2024tyche} format have advanced the state of PBT observability by turning plain textual histograms computed by the libraries into dynamically rendered graphs while enriching the types of feedback libraries can emit during the testing runs.
In Ephyra, observability, particularly through code coverage computed via LLVM coverage counters, is used extensively to guide the evolution of the generator. Generators are not treated as concrete programs that are written, but distributions to be shaped by the reviewers that rely on the observability to understand the underlying tests. We have built prototype interfaces for guiding the generators through the observations collected from the test runs, and we plan to keep developing them in the future.

\section{Ephyra: A Programmable PBT Framework with Synthesizable Components} \label{sec:lib}

Ephyra, the ancient Greek city now known as Corinth, is the home of Sisyphus, a man forever cursed to
carry a boulder up a hill, only to keep going in a metaphorical Möbius strip, never approaching the end. The name of the library is an homage to the most Sisyphean task in modern software development, testing; particularly random testing where we roll not one, but millions of rocks through a variety of software, and Ephyra is the home of all such efforts.
The library follows the programmable PBT tradition of publishing existing property-runners; a simple generational runner, a fuzzing runner and a targeted runner are all within the current implementation. The core innovation, synthesized components, is managed via the accompanying command line interface, and constraints on the generators are denoted via Rust macros as shown in Fig.~\ref{fig:ephyra-generator}.

\begin{figure}[h!]
\centering
\begin{minipage}{0.86\linewidth}
\begin{rustcode}
struct Canonical;
impl Gen for Canonical {
    type Value = String;
    $\overlayfull{generator}{\texttt{fn generate(\&self, g: \&mut Source) -> String}}$ { ... }
    $\overlayfull{shrinker}{\texttt{fn shrink(\&self, s: \&String, visit: ...) -> ControlFlow<()>}}$ { ... }
    $\overlayfull{mutator}{\texttt{fn mutate(\&self, s: \&String, g: \&mut Source) -> String}}$ { ... }
}
// Validity predicate: ada parses the string to itself.
fn is_canonical(s: &String) -> bool {
    canonicalize(s).as_deref() == Some(s.as_str())
}
// Diversity metrics to optimize for the generator distributions
fn canonical_metrics(s: &String) -> Vec<(&'static str, f64)> {
    let b = |x: bool| f64::from(u8::from(x));   // true -> 1.0, false -> 0.0
    vec![("chars", s.chars().count() as f64), ("has query", b(s.contains('?'))), ...]
}

$\overlayfull{generator}{\texttt{\#[ephyra::generator(is\_canonical, metrics = canonical\_metrics)]}}$
fn canonical_urls() -> impl Gen<Value = String> { Canonical }

$\overlayfull{checker}{\texttt{\#[ephyra::test(gen = canonical\_urls(), tests = 3000)]}}$
fn canonical_url_is_reference_canonical(url: String) {
    let r = whatwg_url::run(&Case::parse(url.clone())); // reference implementation
    assert!(r.can_parse);
    assert_eq!(r.href(), Some(url.as_str()));
}
\end{rustcode}
\end{minipage}
\caption{Definition of a generator for the ada-url case study.}
\label{fig:ephyra-generator}
\end{figure}

A validity predicate is an arbitrary Rust function of type \texttt{fn(\&T) -> bool}, generators mark both
their validity predicates as well as the set of metrics to be computed over the distribution of the inputs.
For each annotated generator, \texttt{ephyra check} performs a full Ephyra run over 10{,}000 inputs by default and verifies, per input, that
(1) the generated value satisfies the predicate;
(2) the first 64 shrink candidates differ from the value and satisfy the predicate;
(3) repeatedly taking the first shrink candidate stays valid, never revisits a value, and settles within 1000 steps;
(4) four mutants stay valid; and
(5) every metric is finite.
We use the result of these checks to drive the component synthesis for repairing broken generators, mutators and shrinkers as well as improving accepted ones against the set of computed metrics, complemented by code coverage.

\section{Preliminary Experimental Results} \label{sec:exp}

We applied Ephyra to fourteen real systems in two modes that differ in the steering of the generator synthesis. Appendix~\ref{app:bugs} lists every bug with its upstream report.
In the \emph{open-loop} style, we have run interactive coding-agent sessions that write Ephyra generators, predicates and properties for one target, tune the generators against a metric, run campaigns, and triage what they find.
In the \emph{closed-loop} style, the steering judgement is left up to \texttt{ephyra-agent} with automated patches within the loop to strengthen the ability to find new bugs.

\begin{figure}[h!]
  \centering
  \begin{subfigure}{\linewidth}
    \centering
    \resizebox{0.82\linewidth}{!}{\begin{tikzpicture}[
    font=\sffamily,
    stage/.style={draw, line width=1pt, minimum height=0.8cm, minimum width=1.55cm,
                  align=center, font=\scriptsize\sffamily, inner sep=2pt},
    agent/.style={stage, rounded corners=4pt, draw=mutator, fill=mutator!10},
    ctl/.style={stage, draw=black!70, fill=white},
    gate/.style={font=\tiny\sffamily, text=black!60, align=center},
    flow/.style={->, >={Stealth[length=2mm]}, line width=0.7pt},
    fb/.style={->, >={Stealth[length=2mm]}, line width=0.7pt, dashed},
  ]
  \node[agent] (gen) {Generator\\Synthesis};
  \node[ctl, right=0.6cm of gen] (check) {\texttt{ephyra check}};
  \node[ctl, right=0.9cm of check, minimum width=1.9cm] (run)
       {Campaign runner};
  \node[agent, right=0.95cm of run] (triage) {Triage\\[-1pt]\tiny minimize, bisect, fix};
  \node[stage, below=0.75cm of triage, draw=printer, text=black!70, dashed] (human) {User\\[-1pt]\tiny review, file upstream};

  \draw[flow] (gen) -- (check);
  \draw[flow] (check) -- node[gate, above=1pt] {passes} (run);
  \draw[flow] (run) -- node[gate, above=1pt] {findings} (triage);
  \draw[flow] (triage) -- (human);

  \draw[fb, mutator] (check.south) -- ++(0,-0.5) -| node[gate, pos=0.25, below, text=mutator] {keep / discard} (gen.south);

  \draw[fb, gray] (human.west) -- node[gate, below, pos=0.6] {known bugs,\\fixed target} (human.west -| run.south) -- (run.south);
  \draw[fb, gray] (triage.north) -- ++(0,0.45) -| node[gate, pos=0.25, above] {false positive $\Rightarrow$ narrow the oracle or generator} (gen.north);
\end{tikzpicture}}
    \caption{Open loop: interactive session}
    \label{fig:open-loop}
  \end{subfigure}

  \vspace{0.6em}
  \begin{subfigure}{\linewidth}
    \centering
    \resizebox{0.82\linewidth}{!}{\begin{tikzpicture}[
    font=\sffamily,
    stage/.style={draw, line width=1pt, minimum height=0.8cm, minimum width=1.55cm,
                  align=center, font=\scriptsize\sffamily, inner sep=2pt},
    agent/.style={stage, rounded corners=4pt, draw=mutator, fill=mutator!10},
    ctl/.style={stage, draw=black!70, fill=white},
    gate/.style={font=\tiny\sffamily, text=black!60, align=center},
    flow/.style={->, >={Stealth[length=2mm]}, line width=0.7pt},
    fb/.style={->, >={Stealth[length=2mm]}, line width=0.7pt, dashed},
  ]
  \node[agent] (intake) {Task\\Definition};
  \node[agent, right=0.45cm of intake] (harness) {Test Harness\\Synthesis};
  \node[ctl, right=0.45cm of harness, minimum width=1.9cm] (search)
       {Search\\[-1pt]\tiny 12$\times$30\,min};
  \node[ctl, right=0.9cm of search] (pre) {Pre-triage\\[-1pt]\tiny replay, classify};
  \node[agent, below=0.9cm of pre] (triage) {Triage\\[-1pt]\tiny minimize, fix};
  \node[ctl, left=0.45cm of triage] (verify) {Re-verify\\[-1pt]\tiny repro$\pm$fix};
  \node[agent, left=0.45cm of verify] (writeup) {Write-up\\[-1pt]\tiny dedup, packet};
  \node[stage, left=0.45cm of writeup, draw=printer, text=black!70, dashed] (human) {User\\[-1pt]\tiny validate, file};

  \draw[flow] (intake) -- (harness);
  \draw[flow] (harness) -- node[gate, above=1pt] {} (search);
  \draw[flow] (search) -- node[gate, above=1pt] {findings} (pre);
  \draw[flow] (pre) -- node[gate, right] {new\\bug} (triage);
  \draw[flow] (triage) -- node[gate, above=1pt] {} (verify);
  \draw[flow] (verify) -- (writeup);
  \draw[flow] (writeup) -- (human);

  \draw[fb, black!70] (verify.north) -- node[gate, left] {verified\\patch} (search.south -| verify.north);
  \draw[fb, mutator] (triage.east) -- ++(0.35,0) |- ([yshift=1.05cm]harness.north) node[gate, pos=0.75, above, text=mutator]
       {harness / intended verdict $\Rightarrow$ harness fix, re-gated} -- (harness.north);
  \draw[fb, gray] (pre.north) to[bend right=35] node[gate, above] {drop known signatures} (search.north);
\end{tikzpicture}}
    \caption{Closed loop: the \textsf{ephyra-agent} controller}
    \label{fig:closed-loop}
  \end{subfigure}
  \caption{Two experiment modes: rounded, tinted boxes are LLM agents; square boxes are deterministic tools; the dashed grey box is a user.}
  \label{fig:exp-styles}
\end{figure}
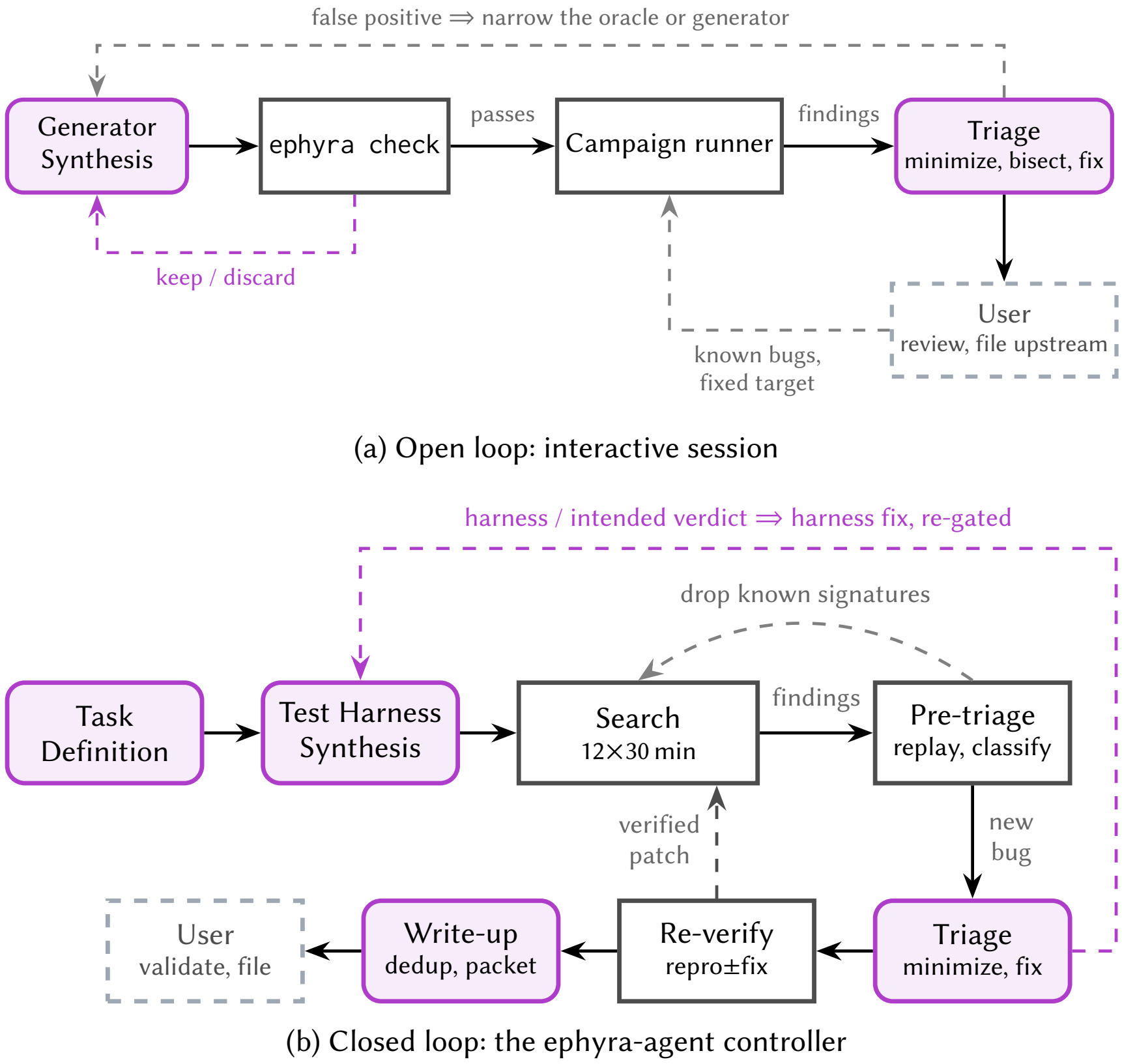

\subsection{Open-Loop Case Studies} \label{sec:exp-open}

We ran open-loop studies on six open-source systems: SQLite (63), Turso (72), PostgreSQL (16), ada-url (12), Cedar (2), and QED (2); 167 confirmed bugs in total. We have used classic properties from the literature such as the metamorphic database testing oracles~\cite{rigger2020tlp,rigger2020norec}, simpler model implementations for model-based testing and differential testing against reference implementations. The ability to generate against arbitrary predicates has given us the opportunity to experiment with new styles of generators, such as directly generating SQLite files from scratch as opposed to the classic approach of running multi-round stateful tests against the database.

\subsection{Closed-Loop Campaign} \label{sec:exp-closed}

The closed-loop experiment ran \textsf{ephyra-agent} (Figure~\ref{fig:closed-loop}) on five Rust and C++ systems in parallel, for about eight hours, one 16-CPU workspace and a \$100 budget each.
Table~\ref{tab:closed-loop} lists, per target, the number of verified bugs, the region coverage of the target's source code, and the agent spend. After the initial campaign, we kept running it informally, finding 72 more bugs; a newer trial on compilers has led to 30 bugs and 11 issues in binaryen, OCaml, and OxCaml.

\vspace{-1mm}

\begin{table}[h!]
  \centering
  \small
  \caption{Closed-loop campaign. Setup (intake and harness), triage and write-up are agent costs in \$.}
  \label{tab:closed-loop}
  \setlength{\tabcolsep}{3pt}
  \begin{tabular}{@{}l l r r r r r r@{}}
    \toprule
    Target & Property & Bugs & Cov. & Setup (\$) & Triage (\$) & Write-up (\$) & Total (\$) \\
    \midrule
    uutils coreutils & differential vs.\ GNU coreutils & 30 & 33.4\% & 11 & 50 & 25 & 87 \\
    brush            & differential vs.\ bash          & 17 & 35.7\% & 13 & 42 & 12 & 67 \\
    gitoxide         & differential vs.\ git           & 14 & 27.3\% & 11 & 24 & 11 & 45 \\
    DuckDB           & TLP, NoREC, optimizer toggles   & 11 & 22.4\% & 20 & 41 & 12 & 74 \\
    DataFusion       & TLP, NoREC, rule toggles        & 8  & 34.9\% & 12 & 37 & 9  & 57 \\
    \midrule
    Total            &                                 & 80 &        & 67 & 195 & 68 & 330 \\
    \bottomrule
  \end{tabular}
\end{table}

\vspace{-4.5mm}

\section{Discussion and Future Work} \label{sec:future}

The main contribution we presented in the experimental results is not unique to Ephyra; it is possible to set up any PBT library with a similar optimization loop, and intuitively expect similar results. We see our experiments as a starting point for agent-authored optimized generators, and Ephyra's design choices as a good set of primitives for such explorations.
Evaluations are immensely more complex to design when LLMs are part of the evaluated system; the classical design of ETNA~\cite{keles2026etnajfp} was not sufficient for our measurements, so we plan to look into existing coding agent evaluation frameworks such as Harbor~\cite{harbor2026} for more controlled experimental settings that only modify the generator, keeping the rest of the system constant to understand the effects in isolation.
One of the crucial problems we encountered was reward hacking. Given a metric, an agent can optimize solely the metric rather than the testing power it stands for. Designing metrics that resist gaming, and detecting the reward hacking remains an open problem for agent-driven generator synthesis.
The future work we imagine is multifaceted: we would like to work on turning Tyche~\cite{goldstein2024tyche} into an interactive tool for shaping generators instead of just understanding them; instead of simply measuring the generators, we would like to prove properties about their soundness and completeness with respect to the desired distribution; we would like to explore the effects of different property-runners in more detail and understand how we can utilize source-level structured mutations more effectively than we do today.
Finally, the optimization architecture we built is a very simple hill-climbing loop. Researchers have shown a variety of different strategies for autonomous optimization loops~\cite{romeraparedes2024funsearch,novikov2025alphaevolve,lange2025shinkaevolve,zhang2025dgm} that we plan to evaluate on generator optimization and adopt when we observe their benefits.

\bibliographystyle{ACM-Reference-Format}
\bibliography{references}


\begin{thebibliography}{43}


\ifx \showCODEN    \undefined \def \showCODEN     #1{\unskip}     \fi
\ifx \showISBNx    \undefined \def \showISBNx     #1{\unskip}     \fi
\ifx \showISBNxiii \undefined \def \showISBNxiii  #1{\unskip}     \fi
\ifx \showISSN     \undefined \def \showISSN      #1{\unskip}     \fi
\ifx \showLCCN     \undefined \def \showLCCN      #1{\unskip}     \fi
\ifx \shownote     \undefined \def \shownote      #1{#1}          \fi
\ifx \showarticletitle \undefined \def \showarticletitle #1{#1}   \fi
\ifx \showURL      \undefined \def \showURL       {\relax}        \fi
\providecommand\bibfield[2]{#2}
\providecommand\bibinfo[2]{#2}
\providecommand\natexlab[1]{#1}
\providecommand\showeprint[2][]{arXiv:#2}
\makeatletter
\@ifundefined{NAT@parse@date}{}{\let\NAT@parse@date@orig\NAT@parse@date}
\@ifundefined{NAT@parse@date}{}{\def\NAT@parse@date#1#2#3#4#5#6@@{\NAT@parse@date@orig#1#2#3#4#5#6@@\def\NAT@tempyear{0000}\def\NAT@tempexlab{{?}}\ifx\NAT@year\NAT@tempyear\ifx\NAT@exlab\NAT@tempexlab\def\NAT@date{[n.\,d.]}\else\edef\NAT@date{[n.\,d.]\NAT@exlab}\fi\fi}}
\makeatother

\bibitem[{Amazon Web Services}(2025)]%
        {kiro}
\bibfield{author}{\bibinfo{person}{{Amazon Web Services}}.} \bibinfo{year}{2025}\natexlab{}.
\newblock \bibinfo{title}{Kiro}.
\newblock \bibinfo{howpublished}{\url{https://kiro.dev/}}.
\newblock
\newblock
\shownote{Accessed 2026-10-20}.


\bibitem[Carlini(2026)]%
        {carlini2026ccompiler}
\bibfield{author}{\bibinfo{person}{Nicholas Carlini}.} \bibinfo{year}{2026}\natexlab{}.
\newblock \bibinfo{title}{Building a {C} compiler with a team of parallel {Claudes}}.
\newblock \bibinfo{howpublished}{\url{https://www.anthropic.com/engineering/building-c-compiler}}.
\newblock
\newblock
\shownote{Anthropic engineering blog, February 5, 2026. Accessed 2026-10-20}.


\bibitem[Chen et~al\mbox{.}(2025)]%
        {chen2025elfuzz}
\bibfield{author}{\bibinfo{person}{Chuyang Chen}, \bibinfo{person}{Brendan Dolan-Gavitt}, {and} \bibinfo{person}{Zhiqiang Lin}.} \bibinfo{year}{2025}\natexlab{}.
\newblock \showarticletitle{{ELFuzz}: Efficient Input Generation via {LLM}-driven Synthesis Over Fuzzer Space}. In \bibinfo{booktitle}{\emph{34th USENIX Security Symposium (USENIX Security 25)}}. \bibinfo{publisher}{USENIX Association}, \bibinfo{address}{Seattle, WA}, \bibinfo{pages}{6279--6298}.
\newblock
\showISBNx{978-1-939133-52-6}
\urldef\tempurl%
\url{https://www.usenix.org/conference/usenixsecurity25/presentation/chen-chuyang}
\showURL{%
\tempurl}


\bibitem[Claessen and Hughes(2000)]%
        {ClaessenH00}
\bibfield{author}{\bibinfo{person}{Koen Claessen} {and} \bibinfo{person}{John Hughes}.} \bibinfo{year}{2000}\natexlab{}.
\newblock \showarticletitle{{QuickCheck}: a lightweight tool for random testing of {Haskell} programs}. In \bibinfo{booktitle}{\emph{Proceedings of the Fifth ACM SIGPLAN International Conference on Functional Programming (ICFP)}}. \bibinfo{publisher}{ACM}, \bibinfo{address}{New York, NY, USA}, \bibinfo{pages}{268--279}.
\newblock
\href{https://doi.org/10.1145/351240.351266}{doi:\nolinkurl{10.1145/351240.351266}}


\bibitem[Deng et~al\mbox{.}(2023)]%
        {deng2023titanfuzz}
\bibfield{author}{\bibinfo{person}{Yinlin Deng}, \bibinfo{person}{Chunqiu~Steven Xia}, \bibinfo{person}{Haoran Peng}, \bibinfo{person}{Chenyuan Yang}, {and} \bibinfo{person}{Lingming Zhang}.} \bibinfo{year}{2023}\natexlab{}.
\newblock \showarticletitle{Large Language Models Are Zero-Shot Fuzzers: Fuzzing Deep-Learning Libraries via Large Language Models}. In \bibinfo{booktitle}{\emph{Proceedings of the 32nd ACM SIGSOFT International Symposium on Software Testing and Analysis (ISSTA)}}. \bibinfo{publisher}{ACM}, \bibinfo{pages}{423--435}.
\newblock
\href{https://doi.org/10.1145/3597926.3598067}{doi:\nolinkurl{10.1145/3597926.3598067}}


\bibitem[Goldstein et~al\mbox{.}(2023)]%
        {goldstein2023reflecting}
\bibfield{author}{\bibinfo{person}{Harrison Goldstein}, \bibinfo{person}{Samantha Frohlich}, \bibinfo{person}{Meng Wang}, {and} \bibinfo{person}{Benjamin~C. Pierce}.} \bibinfo{year}{2023}\natexlab{}.
\newblock \showarticletitle{Reflecting on Random Generation}.
\newblock \bibinfo{journal}{\emph{Proceedings of the ACM on Programming Languages}} \bibinfo{volume}{7}, \bibinfo{number}{ICFP} (\bibinfo{year}{2023}), \bibinfo{pages}{322--355}.
\newblock
\href{https://doi.org/10.1145/3607842}{doi:\nolinkurl{10.1145/3607842}}


\bibitem[Goldstein et~al\mbox{.}(2024)]%
        {goldstein2024tyche}
\bibfield{author}{\bibinfo{person}{Harrison Goldstein}, \bibinfo{person}{Jeffrey Tao}, \bibinfo{person}{Zac Hatfield-Dodds}, \bibinfo{person}{Benjamin~C. Pierce}, {and} \bibinfo{person}{Andrew Head}.} \bibinfo{year}{2024}\natexlab{}.
\newblock \showarticletitle{Tyche: Making Sense of {PBT} Effectiveness}. In \bibinfo{booktitle}{\emph{Proceedings of the 37th Annual ACM Symposium on User Interface Software and Technology (UIST)}}. \bibinfo{publisher}{ACM}, \bibinfo{pages}{1--16}.
\newblock
\href{https://doi.org/10.1145/3654777.3676407}{doi:\nolinkurl{10.1145/3654777.3676407}}


\bibitem[{Harbor Framework Team}(2026)]%
        {harbor2026}
\bibfield{author}{\bibinfo{person}{{Harbor Framework Team}}.} \bibinfo{year}{2026}\natexlab{}.
\newblock \bibinfo{title}{Harbor: A Framework for Evaluating and Optimizing Agents and Models in Container Environments}.
\newblock \bibinfo{howpublished}{\url{https://github.com/harbor-framework/harbor}}.
\newblock
\href{https://doi.org/10.5281/zenodo.20953922}{doi:\nolinkurl{10.5281/zenodo.20953922}}
\newblock
\shownote{Accessed 2026-10-20}.


\bibitem[{Hegel developers}(2026a)]%
        {hegelskill}
\bibfield{author}{\bibinfo{person}{{Hegel developers}}.} \bibinfo{year}{2026}\natexlab{a}.
\newblock \bibinfo{title}{hegel-skill: An Agent Skill for Writing {Hegel} Tests}.
\newblock \bibinfo{howpublished}{\url{https://github.com/hegeldev/hegel-skill}}.
\newblock
\newblock
\shownote{Accessed 2026-10-20}.


\bibitem[{Hegel developers}(2026b)]%
        {hegelzoo}
\bibfield{author}{\bibinfo{person}{{Hegel developers}}.} \bibinfo{year}{2026}\natexlab{b}.
\newblock \bibinfo{title}{hegel-zoo: Property-Based Tests for Open-Source Projects, in Every Language {Hegel} Supports}.
\newblock \bibinfo{howpublished}{\url{https://github.com/hegeldev/hegel-zoo}}.
\newblock
\newblock
\shownote{Accessed 2026-10-20}.


\bibitem[Karpathy(2026)]%
        {karpathy2026autoresearch}
\bibfield{author}{\bibinfo{person}{Andrej Karpathy}.} \bibinfo{year}{2026}\natexlab{}.
\newblock \bibinfo{title}{autoresearch: {AI} Agents Running Research on Single-{GPU} nanochat Training Automatically}.
\newblock \bibinfo{howpublished}{\url{https://github.com/karpathy/autoresearch}}.
\newblock
\newblock
\shownote{Accessed 2026-10-20}.


\bibitem[Keles et~al\mbox{.}(2026a)]%
        {keles2026programmablepropertybasedtesting}
\bibfield{author}{\bibinfo{person}{Alperen Keles}, \bibinfo{person}{Justine Frank}, \bibinfo{person}{Ceren Mert}, \bibinfo{person}{Harrison Goldstein}, {and} \bibinfo{person}{Leonidas Lampropoulos}.} \bibinfo{year}{2026}\natexlab{a}.
\newblock \bibinfo{title}{Programmable Property-Based Testing}.
\newblock
\showeprint[arxiv]{2602.18545}~[cs.SE]
\urldef\tempurl%
\url{https://arxiv.org/abs/2602.18545}
\showURL{%
\tempurl}


\bibitem[Keles et~al\mbox{.}(2026b)]%
        {keles2026evaluatingshrinking}
\bibfield{author}{\bibinfo{person}{Alperen Keles}, \bibinfo{person}{George Miao}, {and} \bibinfo{person}{Leonidas Lampropoulos}.} \bibinfo{year}{2026}\natexlab{b}.
\newblock \bibinfo{title}{Evaluating Shrinking (Experience Report)}.
\newblock
\showeprint[arxiv]{2608.09935}
\urldef\tempurl%
\url{https://arxiv.org/abs/2608.09935}
\showURL{%
\tempurl}
\newblock
\shownote{To appear in Haskell '26: Proceedings of the 19th ACM SIGPLAN International Haskell Symposium}.


\bibitem[Keles et~al\mbox{.}(2026c)]%
        {keles2026etnajfp}
\bibfield{author}{\bibinfo{person}{Alperen Keles}, \bibinfo{person}{Jessica Shi}, \bibinfo{person}{Nikhil Kamath}, \bibinfo{person}{Tin~Nam Liu}, \bibinfo{person}{Ceren Mert}, \bibinfo{person}{Harrison Goldstein}, \bibinfo{person}{Benjamin~C. Pierce}, {and} \bibinfo{person}{Leonidas Lampropoulos}.} \bibinfo{year}{2026}\natexlab{c}.
\newblock \showarticletitle{Etna: An Evaluation Platform for Property-Based Testing}.
\newblock \bibinfo{journal}{\emph{Journal of Functional Programming}}  \bibinfo{volume}{36} (\bibinfo{year}{2026}).
\newblock
\href{https://doi.org/10.46298/jfp.17867}{doi:\nolinkurl{10.46298/jfp.17867}}


\bibitem[Lampropoulos et~al\mbox{.}(2018)]%
        {GeneratingGoodGenerators}
\bibfield{author}{\bibinfo{person}{Leonidas Lampropoulos}, \bibinfo{person}{Zoe Paraskevopoulou}, {and} \bibinfo{person}{Benjamin~C. Pierce}.} \bibinfo{year}{2018}\natexlab{}.
\newblock \showarticletitle{Generating Good Generators for Inductive Relations}.
\newblock \bibinfo{journal}{\emph{Proceedings of the ACM on Programming Languages}} \bibinfo{volume}{2}, \bibinfo{number}{POPL} (\bibinfo{date}{Jan.} \bibinfo{year}{2018}), \bibinfo{numpages}{30}~pages.
\newblock
\href{https://doi.org/10.1145/3158133}{doi:\nolinkurl{10.1145/3158133}}


\bibitem[Lange et~al\mbox{.}(2025)]%
        {lange2025shinkaevolve}
\bibfield{author}{\bibinfo{person}{Robert~Tjarko Lange}, \bibinfo{person}{Yuki Imajuku}, {and} \bibinfo{person}{Edoardo Cetin}.} \bibinfo{year}{2025}\natexlab{}.
\newblock \bibinfo{title}{{ShinkaEvolve}: Towards Open-Ended And Sample-Efficient Program Evolution}.
\newblock
\showeprint[arxiv]{2509.19349}
\urldef\tempurl%
\url{https://arxiv.org/abs/2509.19349}
\showURL{%
\tempurl}


\bibitem[Liu et~al\mbox{.}(2026)]%
        {liu2026aioverlords}
\bibfield{author}{\bibinfo{person}{Shu Liu}, \bibinfo{person}{Soujanya Ponnapalli}, \bibinfo{person}{Shreya Shankar}, \bibinfo{person}{Sepanta Zeighami}, \bibinfo{person}{Alan Zhu}, \bibinfo{person}{Shubham Agarwal}, \bibinfo{person}{Ruiqi Chen}, \bibinfo{person}{Samion Suwito}, \bibinfo{person}{Shuo Yuan}, \bibinfo{person}{Ion Stoica}, \bibinfo{person}{Matei Zaharia}, \bibinfo{person}{Alvin Cheung}, \bibinfo{person}{Natacha Crooks}, \bibinfo{person}{Joseph~E. Gonzalez}, {and} \bibinfo{person}{Aditya~G. Parameswaran}.} \bibinfo{year}{2026}\natexlab{}.
\newblock \showarticletitle{Supporting Our {AI} Overlords: Redesigning Data Systems to be Agent-First}. In \bibinfo{booktitle}{\emph{Conference on Innovative Data Systems Research (CIDR)}}.
\newblock
\urldef\tempurl%
\url{https://www.vldb.org/cidrdb/papers/2026/p32-liu.pdf}
\showURL{%
\tempurl}
\newblock
\shownote{arXiv:2509.00997}.


\bibitem[Maaz et~al\mbox{.}(2025)]%
        {maaz2025agenticpbt}
\bibfield{author}{\bibinfo{person}{Muhammad Maaz}, \bibinfo{person}{Liam DeVoe}, \bibinfo{person}{Zac Hatfield-Dodds}, {and} \bibinfo{person}{Nicholas Carlini}.} \bibinfo{year}{2025}\natexlab{}.
\newblock \bibinfo{title}{Agentic Property-Based Testing: Finding Bugs Across the {Python} Ecosystem}.
\newblock
\showeprint[arxiv]{2510.09907}
\urldef\tempurl%
\url{https://arxiv.org/abs/2510.09907}
\showURL{%
\tempurl}
\newblock
\shownote{NeurIPS 2025 Workshop on Deep Learning for Code}.


\bibitem[MacIver(2026)]%
        {maciver2026hegel}
\bibfield{author}{\bibinfo{person}{David~R. MacIver}.} \bibinfo{year}{2026}\natexlab{}.
\newblock \bibinfo{title}{Hypothesis, Antithesis, synthesis}.
\newblock \bibinfo{howpublished}{\url{https://antithesis.com/blog/2026/hegel/}}.
\newblock
\newblock
\shownote{Antithesis blog, March 24, 2026. Introduces Hegel (\url{https://hegel.dev/}). Accessed 2026-10-20}.


\bibitem[MacIver and Donaldson(2020)]%
        {maciver2020reducer}
\bibfield{author}{\bibinfo{person}{David~R. MacIver} {and} \bibinfo{person}{Alastair~F. Donaldson}.} \bibinfo{year}{2020}\natexlab{}.
\newblock \showarticletitle{Test-Case Reduction via Test-Case Generation: Insights from the {Hypothesis} Reducer (Tool Insights Paper)}. In \bibinfo{booktitle}{\emph{34th European Conference on Object-Oriented Programming (ECOOP 2020)}} \emph{(\bibinfo{series}{LIPIcs}, Vol.~\bibinfo{volume}{166})}. \bibinfo{publisher}{Schloss Dagstuhl -- Leibniz-Zentrum f{\"u}r Informatik}, \bibinfo{pages}{13:1--13:27}.
\newblock
\href{https://doi.org/10.4230/LIPIcs.ECOOP.2020.13}{doi:\nolinkurl{10.4230/LIPIcs.ECOOP.2020.13}}


\bibitem[MacIver et~al\mbox{.}(2019)]%
        {maciver2019hypothesis}
\bibfield{author}{\bibinfo{person}{David~R. MacIver}, \bibinfo{person}{Zac Hatfield-Dodds}, {and} \bibinfo{person}{{Many Contributors}}.} \bibinfo{year}{2019}\natexlab{}.
\newblock \showarticletitle{Hypothesis: A New Approach to Property-Based Testing}.
\newblock \bibinfo{journal}{\emph{Journal of Open Source Software}} \bibinfo{volume}{4}, \bibinfo{number}{43} (\bibinfo{year}{2019}), \bibinfo{pages}{1891}.
\newblock
\href{https://doi.org/10.21105/joss.01891}{doi:\nolinkurl{10.21105/joss.01891}}


\bibitem[Mista et~al\mbox{.}(2018)]%
        {mista2018dragen}
\bibfield{author}{\bibinfo{person}{Agust{\'\i}n Mista}, \bibinfo{person}{Alejandro Russo}, {and} \bibinfo{person}{John Hughes}.} \bibinfo{year}{2018}\natexlab{}.
\newblock \showarticletitle{Branching Processes for {QuickCheck} Generators}. In \bibinfo{booktitle}{\emph{Proceedings of the 11th ACM SIGPLAN International Symposium on Haskell}}. \bibinfo{publisher}{ACM}, \bibinfo{pages}{1--13}.
\newblock
\href{https://doi.org/10.1145/3242744.3242747}{doi:\nolinkurl{10.1145/3242744.3242747}}


\bibitem[Novikov et~al\mbox{.}(2025)]%
        {novikov2025alphaevolve}
\bibfield{author}{\bibinfo{person}{Alexander Novikov}, \bibinfo{person}{Ng{\^a}n V{\~u}}, \bibinfo{person}{Marvin Eisenberger}, \bibinfo{person}{Emilien Dupont}, \bibinfo{person}{Po-Sen Huang}, \bibinfo{person}{Adam~Zsolt Wagner}, \bibinfo{person}{Sergey Shirobokov}, \bibinfo{person}{Borislav Kozlovskii}, \bibinfo{person}{Francisco J.~R. Ruiz}, \bibinfo{person}{Abbas Mehrabian}, \bibinfo{person}{M.~Pawan Kumar}, \bibinfo{person}{Abigail See}, \bibinfo{person}{Swarat Chaudhuri}, \bibinfo{person}{George Holland}, \bibinfo{person}{Alex Davies}, \bibinfo{person}{Sebastian Nowozin}, \bibinfo{person}{Pushmeet Kohli}, {and} \bibinfo{person}{Matej Balog}.} \bibinfo{year}{2025}\natexlab{}.
\newblock \bibinfo{title}{{AlphaEvolve}: A Coding Agent for Scientific and Algorithmic Discovery}.
\newblock
\showeprint[arxiv]{2506.13131}
\urldef\tempurl%
\url{https://arxiv.org/abs/2506.13131}
\showURL{%
\tempurl}


\bibitem[Ouyang et~al\mbox{.}(2025)]%
        {ouyang2025kernelbench}
\bibfield{author}{\bibinfo{person}{Anne Ouyang}, \bibinfo{person}{Simon Guo}, \bibinfo{person}{Simran Arora}, \bibinfo{person}{Alex~L. Zhang}, \bibinfo{person}{William Hu}, \bibinfo{person}{Christopher R{\'e}}, {and} \bibinfo{person}{Azalia Mirhoseini}.} \bibinfo{year}{2025}\natexlab{}.
\newblock \bibinfo{title}{{KernelBench}: Can {LLMs} Write Efficient {GPU} Kernels?}
\newblock
\showeprint[arxiv]{2502.10517}
\urldef\tempurl%
\url{https://arxiv.org/abs/2502.10517}
\showURL{%
\tempurl}


\bibitem[Paraskevopoulou et~al\mbox{.}(2022)]%
        {paraskevopoulou2022computing}
\bibfield{author}{\bibinfo{person}{Zoe Paraskevopoulou}, \bibinfo{person}{Aaron Eline}, {and} \bibinfo{person}{Leonidas Lampropoulos}.} \bibinfo{year}{2022}\natexlab{}.
\newblock \showarticletitle{Computing Correctly with Inductive Relations}. In \bibinfo{booktitle}{\emph{Proceedings of the 43rd ACM SIGPLAN International Conference on Programming Language Design and Implementation (PLDI)}}. \bibinfo{publisher}{ACM}, \bibinfo{pages}{966--980}.
\newblock
\href{https://doi.org/10.1145/3519939.3523707}{doi:\nolinkurl{10.1145/3519939.3523707}}


\bibitem[{Proptest contributors}(2026)]%
        {proptestderive}
\bibfield{author}{\bibinfo{person}{{Proptest contributors}}.} \bibinfo{year}{2026}\natexlab{}.
\newblock \bibinfo{title}{proptest-derive: Custom-Derive for the {Arbitrary} Trait of proptest}.
\newblock \bibinfo{howpublished}{\url{https://docs.rs/proptest-derive/latest/proptest_derive/}}.
\newblock
\newblock
\shownote{Version 0.9.0. Accessed 2026-10-20}.


\bibitem[Rigger and Su(2020a)]%
        {rigger2020norec}
\bibfield{author}{\bibinfo{person}{Manuel Rigger} {and} \bibinfo{person}{Zhendong Su}.} \bibinfo{year}{2020}\natexlab{a}.
\newblock \showarticletitle{Detecting Optimization Bugs in Database Engines via Non-Optimizing Reference Engine Construction}. In \bibinfo{booktitle}{\emph{Proceedings of the 28th ACM Joint Meeting on European Software Engineering Conference and Symposium on the Foundations of Software Engineering (ESEC/FSE)}}. \bibinfo{publisher}{ACM}, \bibinfo{pages}{1140--1152}.
\newblock
\href{https://doi.org/10.1145/3368089.3409710}{doi:\nolinkurl{10.1145/3368089.3409710}}


\bibitem[Rigger and Su(2020b)]%
        {rigger2020tlp}
\bibfield{author}{\bibinfo{person}{Manuel Rigger} {and} \bibinfo{person}{Zhendong Su}.} \bibinfo{year}{2020}\natexlab{b}.
\newblock \showarticletitle{Finding Bugs in Database Systems via Query Partitioning}.
\newblock \bibinfo{journal}{\emph{Proceedings of the ACM on Programming Languages}} \bibinfo{volume}{4}, \bibinfo{number}{OOPSLA} (\bibinfo{year}{2020}), \bibinfo{pages}{1--30}.
\newblock
\href{https://doi.org/10.1145/3428279}{doi:\nolinkurl{10.1145/3428279}}


\bibitem[Romera-Paredes et~al\mbox{.}(2023)]%
        {romeraparedes2024funsearch}
\bibfield{author}{\bibinfo{person}{Bernardino Romera-Paredes}, \bibinfo{person}{Mohammadamin Barekatain}, \bibinfo{person}{Alexander Novikov}, \bibinfo{person}{Matej Balog}, \bibinfo{person}{M.~Pawan Kumar}, \bibinfo{person}{Emilien Dupont}, \bibinfo{person}{Francisco J.~R. Ruiz}, \bibinfo{person}{Jordan~S. Ellenberg}, \bibinfo{person}{Pengming Wang}, \bibinfo{person}{Omar Fawzi}, \bibinfo{person}{Pushmeet Kohli}, {and} \bibinfo{person}{Alhussein Fawzi}.} \bibinfo{year}{2023}\natexlab{}.
\newblock \showarticletitle{Mathematical discoveries from program search with large language models}.
\newblock \bibinfo{journal}{\emph{Nature}} \bibinfo{volume}{625}, \bibinfo{number}{7995} (\bibinfo{year}{2023}), \bibinfo{pages}{468--475}.
\newblock
\href{https://doi.org/10.1038/s41586-023-06924-6}{doi:\nolinkurl{10.1038/s41586-023-06924-6}}


\bibitem[Ronacher(2025)]%
        {ronacher2025buildityourself}
\bibfield{author}{\bibinfo{person}{Armin Ronacher}.} \bibinfo{year}{2025}\natexlab{}.
\newblock \bibinfo{title}{Build It Yourself}.
\newblock \bibinfo{howpublished}{\url{https://lucumr.pocoo.org/2025/1/24/build-it-yourself/}}.
\newblock
\newblock
\shownote{Blog post}.


\bibitem[Solar-Lezama et~al\mbox{.}(2006)]%
        {solarlezama2006sketching}
\bibfield{author}{\bibinfo{person}{Armando Solar-Lezama}, \bibinfo{person}{Liviu Tancau}, \bibinfo{person}{Rastislav Bod{\'\i}k}, \bibinfo{person}{Sanjit Seshia}, {and} \bibinfo{person}{Vijay Saraswat}.} \bibinfo{year}{2006}\natexlab{}.
\newblock \showarticletitle{Combinatorial Sketching for Finite Programs}. In \bibinfo{booktitle}{\emph{Proceedings of the 12th International Conference on Architectural Support for Programming Languages and Operating Systems (ASPLOS)}}. \bibinfo{publisher}{ACM}, \bibinfo{pages}{404--415}.
\newblock
\href{https://doi.org/10.1145/1168857.1168907}{doi:\nolinkurl{10.1145/1168857.1168907}}


\bibitem[Sumner(2026)]%
        {sumner2026bunrust}
\bibfield{author}{\bibinfo{person}{Jarred Sumner}.} \bibinfo{year}{2026}\natexlab{}.
\newblock \bibinfo{title}{Rewriting {Bun} in {Rust}}.
\newblock \bibinfo{howpublished}{\url{https://bun.com/blog/bun-in-rust}}.
\newblock
\newblock
\shownote{Bun blog, July 8, 2026. Accessed 2026-10-20}.


\bibitem[Tomlinson(2026)]%
        {tomlinson2026libraries}
\bibfield{author}{\bibinfo{person}{Jacob Tomlinson}.} \bibinfo{year}{2026}\natexlab{}.
\newblock \bibinfo{title}{Open Source Libraries Are Evolving into Reference Implementations}.
\newblock \bibinfo{howpublished}{\url{https://jacobtomlinson.dev/posts/2026/open-source-libraries-are-evolving-into-reference-implementations/}}.
\newblock
\newblock
\shownote{Blog post, August 11, 2026. Accessed 2026-10-20}.


\bibitem[{Trail of Bits}(2026)]%
        {trailofbitspbtskill}
\bibfield{author}{\bibinfo{person}{{Trail of Bits}}.} \bibinfo{year}{2026}\natexlab{}.
\newblock \bibinfo{title}{Property-Based Testing Skill, {Trail of Bits} Skills}.
\newblock \bibinfo{howpublished}{\url{https://github.com/trailofbits/skills/tree/main/plugins/property-based-testing}}.
\newblock
\newblock
\shownote{Accessed 2026-10-20}.


\bibitem[Vikram et~al\mbox{.}(2024)]%
        {vikram2024llmpbt}
\bibfield{author}{\bibinfo{person}{Vasudev Vikram}, \bibinfo{person}{Caroline Lemieux}, \bibinfo{person}{Joshua Sunshine}, {and} \bibinfo{person}{Rohan Padhye}.} \bibinfo{year}{2024}\natexlab{}.
\newblock \bibinfo{title}{Can Large Language Models Write Good Property-Based Tests?}
\newblock
\showeprint[arxiv]{2307.04346}~[cs.SE]
\urldef\tempurl%
\url{https://arxiv.org/abs/2307.04346}
\showURL{%
\tempurl}


\bibitem[Vikram and Padhye(2026)]%
        {vikram2026gentoo}
\bibfield{author}{\bibinfo{person}{Vasudev Vikram} {and} \bibinfo{person}{Rohan Padhye}.} \bibinfo{year}{2026}\natexlab{}.
\newblock \bibinfo{title}{Fuzzing with Agents? Generators Are All You Need}.
\newblock
\showeprint[arxiv]{2604.01442}
\urldef\tempurl%
\url{https://arxiv.org/abs/2604.01442}
\showURL{%
\tempurl}


\bibitem[Willison(2026)]%
        {willison2026porting}
\bibfield{author}{\bibinfo{person}{Simon Willison}.} \bibinfo{year}{2026}\natexlab{}.
\newblock \bibinfo{title}{My answers to the questions I posed about porting open source code with {LLMs}}.
\newblock \bibinfo{howpublished}{\url{https://simonwillison.net/2026/Jan/11/answers/}}.
\newblock
\newblock
\shownote{Blog post}.


\bibitem[Xia et~al\mbox{.}(2024)]%
        {xia2024fuzz4all}
\bibfield{author}{\bibinfo{person}{Chunqiu~Steven Xia}, \bibinfo{person}{Matteo Paltenghi}, \bibinfo{person}{Jia Le~Tian}, \bibinfo{person}{Michael Pradel}, {and} \bibinfo{person}{Lingming Zhang}.} \bibinfo{year}{2024}\natexlab{}.
\newblock \showarticletitle{{Fuzz4All}: Universal Fuzzing with Large Language Models}. In \bibinfo{booktitle}{\emph{Proceedings of the IEEE/ACM 46th International Conference on Software Engineering (ICSE)}}. \bibinfo{publisher}{ACM}, \bibinfo{pages}{1--13}.
\newblock
\href{https://doi.org/10.1145/3597503.3639121}{doi:\nolinkurl{10.1145/3597503.3639121}}


\bibitem[Yang et~al\mbox{.}(2024)]%
        {yang2024whitefox}
\bibfield{author}{\bibinfo{person}{Chenyuan Yang}, \bibinfo{person}{Yinlin Deng}, \bibinfo{person}{Runyu Lu}, \bibinfo{person}{Jiayi Yao}, \bibinfo{person}{Jiawei Liu}, \bibinfo{person}{Reyhaneh Jabbarvand}, {and} \bibinfo{person}{Lingming Zhang}.} \bibinfo{year}{2024}\natexlab{}.
\newblock \showarticletitle{{WhiteFox}: White-Box Compiler Fuzzing Empowered by Large Language Models}.
\newblock \bibinfo{journal}{\emph{Proceedings of the ACM on Programming Languages}} \bibinfo{volume}{8}, \bibinfo{number}{OOPSLA2} (\bibinfo{year}{2024}), \bibinfo{pages}{709--735}.
\newblock
\href{https://doi.org/10.1145/3689736}{doi:\nolinkurl{10.1145/3689736}}


\bibitem[Yang et~al\mbox{.}(2026)]%
        {yang2026programbench}
\bibfield{author}{\bibinfo{person}{John Yang}, \bibinfo{person}{Kilian Lieret}, \bibinfo{person}{Jeffrey Ma}, \bibinfo{person}{Parth Thakkar}, \bibinfo{person}{Dmitrii Pedchenko}, \bibinfo{person}{Sten Sootla}, \bibinfo{person}{Emily McMilin}, \bibinfo{person}{Pengcheng Yin}, \bibinfo{person}{Rui Hou}, \bibinfo{person}{Gabriel Synnaeve}, \bibinfo{person}{Diyi Yang}, {and} \bibinfo{person}{Ofir Press}.} \bibinfo{year}{2026}\natexlab{}.
\newblock \bibinfo{title}{{ProgramBench}: Can Language Models Rebuild Programs From Scratch?}
\newblock
\showeprint[arxiv]{2605.03546}
\urldef\tempurl%
\url{https://arxiv.org/abs/2605.03546}
\showURL{%
\tempurl}


\bibitem[Zhang et~al\mbox{.}(2025a)]%
        {zhang2025dgm}
\bibfield{author}{\bibinfo{person}{Jenny Zhang}, \bibinfo{person}{Shengran Hu}, \bibinfo{person}{Cong Lu}, \bibinfo{person}{Robert Lange}, {and} \bibinfo{person}{Jeff Clune}.} \bibinfo{year}{2025}\natexlab{a}.
\newblock \bibinfo{title}{Darwin {G}{\"o}del Machine: Open-Ended Evolution of Self-Improving Agents}.
\newblock
\showeprint[arxiv]{2505.22954}
\urldef\tempurl%
\url{https://arxiv.org/abs/2505.22954}
\showURL{%
\tempurl}


\bibitem[Zhang et~al\mbox{.}(2025b)]%
        {zhang2025g2fuzz}
\bibfield{author}{\bibinfo{person}{Kunpeng Zhang}, \bibinfo{person}{Zongjie Li}, \bibinfo{person}{Daoyuan Wu}, \bibinfo{person}{Shuai Wang}, {and} \bibinfo{person}{Xin Xia}.} \bibinfo{year}{2025}\natexlab{b}.
\newblock \bibinfo{title}{Low-Cost and Comprehensive Non-textual Input Fuzzing with {LLM}-Synthesized Input Generators}.
\newblock
\showeprint[arxiv]{2501.19282}
\urldef\tempurl%
\url{https://arxiv.org/abs/2501.19282}
\showURL{%
\tempurl}
\newblock
\shownote{34th USENIX Security Symposium}.


\bibitem[Zhang et~al\mbox{.}(2024)]%
        {zhang2024lpr}
\bibfield{author}{\bibinfo{person}{Mengxiao Zhang}, \bibinfo{person}{Yongqiang Tian}, \bibinfo{person}{Zhenyang Xu}, \bibinfo{person}{Yiwen Dong}, \bibinfo{person}{Shin~Hwei Tan}, {and} \bibinfo{person}{Chengnian Sun}.} \bibinfo{year}{2024}\natexlab{}.
\newblock \showarticletitle{{LPR}: Large Language Models-Aided Program Reduction}. In \bibinfo{booktitle}{\emph{Proceedings of the 33rd ACM SIGSOFT International Symposium on Software Testing and Analysis (ISSTA)}}. \bibinfo{publisher}{ACM}, \bibinfo{pages}{261--273}.
\newblock
\href{https://doi.org/10.1145/3650212.3652126}{doi:\nolinkurl{10.1145/3650212.3652126}}


\end{thebibliography}

\appendix
\section{Bug Reports} \label{app:bugs}

Table~\ref{tab:bugs} lists the 349 bugs found in the case studies of \S\ref{sec:exp}: 167 in the open-loop studies, 80 in the closed-loop campaign, 72 in a later closed-loop run, and 30 in a later trial on compilers. \emph{Duplicate} means an existing upstream issue or pull request already covers the bug; \emph{won't report} means we decided not to report the bug; \emph{not yet reported} means its write-up is ready but not filed; \emph{declined} means upstream closed our report as not planned.

{\footnotesize
\begin{longtable}{@{}p{0.6\linewidth} l l@{}}
\caption{Bugs found by the case studies, with their upstream reports.}\label{tab:bugs}\\
\toprule Bug & Report & Status \\ \midrule \endfirsthead
\toprule Bug & Report & Status \\ \midrule \endhead
\bottomrule \endfoot
\multicolumn{3}{@{}l}{\textbf{SQLite} (open loop, 63)} \\*
{}ROLLBACK TO after DROP TABLE/INDEX commits a stale root page (incremental vacuum) & \href{https://sqlite.org/bugs/forumpost/3bc3c8adfa}{forum 3bc3c8adfa} & fixed \\
{}DELETE leaves an unaccounted byte on an interior page (WITHOUT ROWID, 3-byte key) & -- & fixed \\
{}x IN (SELECT ...) in HAVING or the result converts the GROUP BY term x & -- & fixed \\
{}Blob over 4000 bytes read from overflow pages is converted as UTF-8 in a UTF-16 database & \href{https://sqlite.org/bugs/forumpost/48c64d8bab}{forum 48c64d8bab} & fixed \\
{}LEFT JOIN with b = NULL AND c IN (...) on index (b, c) also returns a matched row & \href{https://sqlite.org/bugs/forumpost/78bc86a164}{forum 78bc86a164} & fixed \\
{}SELECT * over a USING join left of a RIGHT/FULL JOIN fails with ambiguous column name & \href{https://sqlite.org/bugs/forumpost/8be1215e2c}{forum 8be1215e2c} & reported \\
{}Flattened UNION ALL subquery with a RIGHT/FULL JOIN after its first SELECT loses WHERE terms & \href{https://sqlite.org/bugs/forumpost/d098617140}{forum d098617140} & fixed \\
{}Window query drops the left row of a correlated LEFT JOIN whose ON has a bare outer column & \href{https://sqlite.org/bugs/forumpost/7f4ae01b58}{forum 7f4ae01b58} & fixed \\
{}UNION ALL subquery/CTE with inner join left of a RIGHT/FULL JOIN returns only its last arm & \href{https://sqlite.org/bugs/forumpost/50dd78a44b}{forum 50dd78a44b} & fixed \\
{}GROUP BY/ORDER BY over a subquery ordered the other direction skips the outer sort & \href{https://sqlite.org/bugs/forumpost/5cd4b3b27a}{forum 5cd4b3b27a} & fixed \\
{}json\_valid() returns 0 for canonical JSON made from JSON5 input (stale cached parse) & \href{https://sqlite.org/bugs/forumpost/1339f93b28}{forum 1339f93b28} & fixed \\
{}Join drops rows with an expression index over a column pinned by col = constant & \href{https://sqlite.org/bugs/forumpost/529732a774}{forum 529732a774} & fixed \\
{}MULTI-INDEX OR over a partial index WHERE a=const fails with internal query planner error & \href{https://sqlite.org/bugs/forumpost/3ede79ce42}{forum 3ede79ce42} & fixed \\
{}Partial index WHERE col = constant makes col read back as the constant with col's affinity & \href{https://sqlite.org/bugs/forumpost/35992eec46}{forum 35992eec46} & partly fixed \\
{}Bloom filter before an INTEGER PRIMARY KEY lookup drops the REAL join key -2\textasciicircum{}63 & \href{https://sqlite.org/bugs/forumpost/403ca95409}{forum 403ca95409} & fixed \\
{}Constant propagation leaves REAL -2\textasciicircum{}63 a REAL and misses an INTEGER -2\textasciicircum{}63 row & \href{https://sqlite.org/bugs/forumpost/9bbaa34eea}{forum 9bbaa34eea} & partly fixed \\
{}sum(b ORDER BY a) aggregates the ORDER BY term instead of the argument & \href{https://sqlite.org/bugs/forumpost/b53a65d1fc}{forum b53a65d1fc} & fixed \\
{}TEXT over 4000 bytes read from overflow pages loses a leading BOM in a UTF-16 database & \href{https://sqlite.org/bugs/forumpost/36a73e8c8a}{forum 36a73e8c8a} & fixed \\
{}JoinMiss optimization reuses a pruned automatic index across correlated subquery runs & \href{https://sqlite.org/bugs/forumpost/8e5c07be86}{forum 8e5c07be86} & fixed \\
{}LIKE/GLOB optimization's two-range index scan is treated as ordered under IN/skip-scan & \href{https://sqlite.org/bugs/forumpost/8cbaf327c7}{forum 8cbaf327c7} & fixed \\
{}Planner ignores NOT INDEXED on WITHOUT ROWID tables (plan only) & \href{https://sqlite.org/bugs/forumpost/cd41fceaa0}{forum cd41fceaa0} & fixed \\
{}x IN (SELECT y COLLATE NOCASE ...) seeks a BINARY index and misses rows & \href{https://sqlite.org/bugs/forumpost/f505c400cf}{forum f505c400cf} & fixed \\
{}ADD COLUMN blob DEFAULT is read as UTF-8 when cast to text in a UTF-16 database & \href{https://sqlite.org/bugs/forumpost/a9c654a346}{forum a9c654a346} & fixed \\
{}Index on an untyped VIRTUAL generated column over REAL returns integral REALs as INTEGER & \href{https://sqlite.org/bugs/forumpost/6973ad169e}{forum 6973ad169e} & fixed \\
{}WITHOUT ROWID ORDER BY added column NULLS LAST with PK IN: malformed error or wrong rows & \href{https://sqlite.org/bugs/forumpost/eee345f931}{forum eee345f931} & fixed \\
{}group\_concat() window drops a separator after an inverse step leaves the text empty & \href{https://sqlite.org/bugs/forumpost/b02f0a01e0}{forum b02f0a01e0} & fixed \\
{}WITHOUT ROWID row-value range into an added column seeks the PK: malformed or missing rows & \href{https://sqlite.org/bugs/forumpost/eee345f931}{forum eee345f931 (reply)} & reported \\
{}concat()/concat\_ws() take a BLOB's size before UTF-8 conversion in a UTF-16 database & \href{https://sqlite.org/bugs/forumpost/c621e963a9}{forum c621e963a9} & duplicate \\
{}Row-value IN list driving a join reuses the full IN's ephemeral table, repeating rows & \href{https://sqlite.org/bugs/forumpost/764df92da2}{forum 764df92da2} & fixed \\
{}Row-value IN (SELECT a, b) over WITHOUT ROWID with added column b probes the PK: malformed & \href{https://sqlite.org/bugs/forumpost/eee345f931}{forum eee345f931 (reply)} & reported \\
{}Compound with ORDER BY COLLATE applies the last SELECT's GROUP BY/HAVING/DISTINCT to all & \href{https://sqlite.org/bugs/forumpost/3859ed06b6}{forum 3859ed06b6} & fixed \\
{}OR in a LEFT JOIN's ON with the no-op join table omitted: wrong rows or SIGSEGV & \href{https://sqlite.org/bugs/forumpost/16c7c8892b}{forum 16c7c8892b} & fixed \\
{}Btree delete keeps the deleted cell's cached CellInfo; a later write reports malformed & \href{https://sqlite.org/bugs/forumpost/3e17701ba5}{forum 3e17701ba5} & fixed \\
{}Compound merge removes duplicates under one operand's collation, not the compound's & \href{https://sqlite.org/bugs/forumpost/a3d1c48f31}{forum a3d1c48f31} & fixed \\
{}RENAME COLUMN of a TEMP table rewrites temp triggers and views twice, breaking them & \href{https://sqlite.org/bugs/forumpost/c75ba2bcd0}{forum c75ba2bcd0} & fixed \\
{}FTS5 ft = 'q': constant propagation breaks auxiliary functions (no such cursor) & \href{https://sqlite.org/bugs/forumpost/a27cfbee6c}{forum a27cfbee6c} & reported \\
{}FTS5 MATCH with rowid = non-integer returns every matching row & \href{https://sqlite.org/bugs/forumpost/422031d9f4}{forum 422031d9f4} & fixed \\
{}Virtual table created in a savepoint skips xRollbackTo; FTS transaction then fails & \href{https://sqlite.org/bugs/forumpost/9454b36e0a}{forum 9454b36e0a} & fixed \\
{}FTS5 detail=none/column tokenizes UNINDEXED columns; snippet() reports malformed & \href{https://sqlite.org/bugs/forumpost/a3eee51a91}{forum a3eee51a91} & fixed \\
{}Geopoly rowid = non-integer value finds the truncated rowid; UPDATE/DELETE hit the wrong row & \href{https://sqlite.org/bugs/forumpost/6d40716f60}{forum 6d40716f60} & fixed \\
{}FTS5 locale=1, contentless\_unindexed=1 with an UNINDEXED column: every INSERT is malformed & \href{https://sqlite.org/bugs/forumpost/c7ab2c9250}{forum c7ab2c9250} & reported \\
{}FTS5 tokendata=1 query for a term missing from a segment reads another page: corruption error & \href{https://sqlite.org/bugs/forumpost/8a9e1dca7e}{forum 8a9e1dca7e} & fixed \\
{}FTS5 contentless\_unindexed table: DROP leaves \%\_content behind and RENAME skips it & \href{https://sqlite.org/bugs/forumpost/87d4bdefc0}{forum 87d4bdefc0} & fixed \\
{}Geopoly UPDATE of the rowid without \_shape fails with an invalid polygon error & \href{https://sqlite.org/bugs/forumpost/1b9d323951}{forum 1b9d323951} & fixed \\
{}R-Tree UPDATE SET rowid / INSERT with rowid ignore the new rowid & \href{https://sqlite.org/bugs/forumpost/26da7cd9e2}{forum 26da7cd9e2} & fixed \\
{}FTS5 secure-delete: stale doclist-index jump returns rows of the next term\textsuperscript{\dag} & -- & fixed \\
{}FTS5 detail=none: secure-delete of a rowid replaced with secure-delete off corrupts the index & \href{https://sqlite.org/bugs/forumpost/e380b6876c}{forum e380b6876c} & reported \\
{}fts5vocab term range constraints and ORDER BY term use UTF-8 order in a UTF-16 database & \href{https://sqlite.org/bugs/forumpost/b0c03988e8}{forum b0c03988e8} & reported \\
{}Compound rows change with ORDER BY when a later member sets the collation (since 3.53) & \href{https://sqlite.org/bugs/forumpost/2fb4734a27}{forum 2fb4734a27} & reported \\
{}FTS5 secure-delete: a merge drops a delete marker, so a later secure delete revives old entry\textsuperscript{\dag} & \href{https://sqlite.org/bugs/forumpost/737cb495ea}{forum 737cb495ea} & reported \\
{}DROP CONSTRAINT of a foreign key with SET DEFAULT/NOT DEFERRABLE/MATCH fails or leaves part\textsuperscript{\dag} & \href{https://sqlite.org/bugs/forumpost/5e80e94512}{forum 5e80e94512} & fixed \\
{}DROP NOT NULL removes NOT DEFERRABLE instead, or drops a DEFERRABLE clause after NOT NULL\textsuperscript{\dag} & \href{https://sqlite.org/bugs/forumpost/28af38d7c3}{forum 28af38d7c3} & fixed \\
{}FTS5: phrases of a NEAR group that does not match a row report stale instances for it & \href{https://sqlite.org/bugs/forumpost/9fa08e2dbd}{forum 9fa08e2dbd} & reported \\
{}FTS5: wrong instances below a non-matching AND, and right of a NOT whose left side is empty\textsuperscript{\dag} & \href{https://sqlite.org/bugs/forumpost/5cffabd8e9}{forum 5cffabd8e9} & reported \\
{}fts5vocab rowids restart in every scan, so MULTI-INDEX OR and RIGHT JOIN return wrong rows & \href{https://sqlite.org/bugs/forumpost/d29f636f37}{forum d29f636f37} & reported \\
{}RENAME COLUMN fails on an INTEGER PRIMARY KEY column named in a UNIQUE table constraint & \href{https://sqlite.org/bugs/forumpost/b088eb5b02}{forum b088eb5b02} & reported \\
{}MULTI-INDEX OR on a WITHOUT ROWID table misses rows when an OR term uses OP\_SeekScan & \href{https://sqlite.org/bugs/forumpost/f554fe3656}{forum f554fe3656} & fixed \\
{}FTS5 trigram: LIKE and GLOB on an UNINDEXED column return no rows & \href{https://sqlite.org/bugs/forumpost/be50ab5692}{forum be50ab5692} & reported \\
{}WHERE OR on a table left of a RIGHT JOIN drives a MULTI-INDEX OR and is never tested & -- & won't report \\
{}FTS5 tokendata=1: prefix query then term query in an open transaction reads freed memory & \href{https://sqlite.org/bugs/forumpost/5c53f75fbb}{forum 5c53f75fbb} & fixed \\
{}FTS5 trigram LIKE pushdown reads one byte past the pattern (sqlite3Fts5ExprPattern)\textsuperscript{\dag} & \href{https://sqlite.org/bugs/forumpost/402c7177cb}{forum 402c7177cb} & fixed \\
{}In a UTF-16 database a zeroblob() used as text is converted from UTF-8, so CAST(zeroblob(4) AS TEXT) has 8 bytes and depends on the plan & \href{https://sqlite.org/bugs/forumpost/420911fdac}{forum 420911fdac} & fixed \\
{}UPDATE re-applies affinity to unassigned columns: REAL -2\textasciicircum{}63 becomes INTEGER in the table but not in untouched indexes (trunk regression) & -- & won't report \\
\addlinespace
\multicolumn{3}{@{}l}{\textbf{Turso} (open loop, 72)} \\*
{}X -> X / X ->> X with identical operands panics (index out of bounds) & \href{https://github.com/tursodatabase/turso/issues/9836}{\#9836} & reported \\
{}Date/time modifiers accept trailing text & \href{https://github.com/tursodatabase/turso/issues/9837}{\#9837} & fixed \\
{}json\_each/json\_tree with a NULL root path errors instead of returning no rows & \href{https://github.com/tursodatabase/turso/issues/9838}{\#9838} & reported \\
{}json\_each/json\_tree error on an integer beyond i64 instead of returning a real & \href{https://github.com/tursodatabase/turso/issues/9839}{\#9839} & fixed \\
{}Negative-zero REAL becomes -0.0 in JSON; SQLite writes 0.0 & \href{https://github.com/tursodatabase/turso/issues/9840}{\#9840} & reported \\
{}json\_error\_position at end of input after whitespace: internal error or off by one & \href{https://github.com/tursodatabase/turso/issues/9841}{\#9841} & fixed \\
{}Unused CTE is resolved and its errors reported when the statement has any subquery & \href{https://github.com/tursodatabase/turso/issues/9842}{\#9842} & fixed \\
{}json\_patch with a repeated key (object then null) gives malformed JSONB or wrong text & \href{https://github.com/tursodatabase/turso/issues/9843}{\#9843} & reported \\
{}json\_set/jsonb\_set/json\_remove mishandle NULL results, inputs and paths & \href{https://github.com/tursodatabase/turso/issues/9844}{\#9844} & reported \\
{}->/->> abbreviated, 32-bit integer and BLOB paths return NULL & \href{https://github.com/tursodatabase/turso/issues/9845}{\#9845} & reported \\
{}RIGHT/FULL JOIN with a WHERE subquery over a table-valued function panics & \href{https://github.com/tursodatabase/turso/issues/9847}{\#9847} & fixed \\
{}MATERIALIZED CTE with a self-referencing subquery in a nested query: internal error & \href{https://github.com/tursodatabase/turso/issues/9848}{\#9848} & fixed \\
{}Text-to-real parsing for REAL affinity and sum/avg differs from CAST beyond exponent \textasciitilde{}99 & \href{https://github.com/tursodatabase/turso/issues/9849}{\#9849} & reported \\
{}RIGHT JOIN then LEFT JOIN whose ON subquery reads the left table: panic or wrong rows & \href{https://github.com/tursodatabase/turso/issues/9850}{\#9850} & reported \\
{}Subqueries and constants hoisted before the loop raise errors SQLite never hits & \href{https://github.com/tursodatabase/turso/issues/9851}{\#9851} & reported \\
{}FULL/RIGHT JOIN onto a WITHOUT ROWID table with a secondary index panics & \href{https://github.com/tursodatabase/turso/issues/9852}{\#9852} & fixed \\
{}HAVING terms on GROUP BY columns are not applied before aggregating & \href{https://github.com/tursodatabase/turso/issues/9880}{\#9880} & fixed \\
{}Hash join returns a whole-number REAL stored as an integer as an INTEGER & \href{https://github.com/tursodatabase/turso/issues/9881}{\#9881} & fixed \\
{}MATERIALIZED CTE first filled inside a nested subquery's loop panics & \href{https://github.com/tursodatabase/turso/issues/9882}{\#9882} & fixed \\
{}min/max window with a frame and outer ORDER BY rowid fails with table not found & \href{https://github.com/tursodatabase/turso/issues/9883}{\#9883} & fixed \\
{}strftime() ignores the julian-day range check & \href{https://github.com/tursodatabase/turso/issues/9884}{\#9884} & fixed \\
{}json\_pretty of an empty object/array adds a newline & \href{https://github.com/tursodatabase/turso/issues/9885}{\#9885} & reported \\
{}timediff skips day>28 normalization & \href{https://github.com/tursodatabase/turso/issues/9886}{\#9886} & fixed \\
{}HAVING predicate that is a GROUP BY term reads the column after the scan & \href{https://github.com/tursodatabase/turso/issues/9887}{\#9887} & fixed \\
{}strftime \%J / REAL formatting prints 17 digits & \href{https://github.com/tursodatabase/turso/issues/9888}{\#9888} & reported \\
{}-0.0 in a REAL column keeps its sign & \href{https://github.com/tursodatabase/turso/issues/9889}{\#9889} & fixed \\
{}+HH:MM modifier with hour 24 adds a day & \href{https://github.com/tursodatabase/turso/issues/9921}{\#9921} & fixed \\
{}json\_insert at the root path replaces the document & \href{https://github.com/tursodatabase/turso/issues/9924}{\#9924} & fixed \\
{}-9223372036854775808 and -9.223372036854776e18 treated as the same expression & \href{https://github.com/tursodatabase/turso/issues/9925}{\#9925} & fixed \\
{}BETWEEN on a partial index's column panics (covering index misjudged) & \href{https://github.com/tursodatabase/turso/issues/9926}{\#9926} & fixed \\
{}JSON infinity rendering by global string replace corrupts JSON strings & \href{https://github.com/tursodatabase/turso/issues/9927}{\#9927} & fixed \\
{}UPDATE through a covering index with a rowid-reading partial index: panic or duplicate rowid & \href{https://github.com/tursodatabase/turso/issues/9932}{\#9932} & fixed \\
{}UPDATE with a BEFORE UPDATE trigger builds an index key without REAL affinity: corrupt index & \href{https://github.com/tursodatabase/turso/issues/9933}{\#9933} & fixed \\
{}Partial index + join panics: column not stored in the index & \href{https://github.com/tursodatabase/turso/issues/9934}{\#9934} & fixed \\
{}Date/time functions don't stop at an embedded NUL & \href{https://github.com/tursodatabase/turso/issues/9935}{\#9935} & fixed \\
{}JSON subtype lost when a result column contains a GROUP BY expression & \href{https://github.com/tursodatabase/turso/issues/9966}{\#9966} & fixed \\
{}JSON cache treats a REAL 1.0 as the INTEGER 1 seen just before & \href{https://github.com/tursodatabase/turso/issues/9967}{\#9967} & fixed \\
{}Date/time functions reject a trailing vertical tab after the time value & \href{https://github.com/tursodatabase/turso/issues/9968}{\#9968} & fixed \\
{}strftime() prints an overflowing day as is instead of normalizing it & \href{https://github.com/tursodatabase/turso/issues/9969}{\#9969} & fixed \\
{}CREATE INDEX rejects double-quoted string literals SQLite accepts & \href{https://github.com/tursodatabase/turso/issues/10337}{\#10337} & reported \\
{}likelihood() rejects a float literal with an exponent (625e-4) or in parentheses & \href{https://github.com/tursodatabase/turso/issues/10338}{\#10338} & reported \\
{}pragma\_quick\_check / pragma\_integrity\_check name their column message & \href{https://github.com/tursodatabase/turso/issues/10339}{\#10339} & reported \\
{}ANALYZE cannot reach the temp schema or TEMP tables and indexes & \href{https://github.com/tursodatabase/turso/issues/10340}{\#10340} & fixed \\
{}VACUUM INTO ':memory:' writes a file named :memory:; VACUUM INTO '' rejected & \href{https://github.com/tursodatabase/turso/issues/10341}{\#10341} & fixed \\
{}json\_each/json\_tree declare column types, so comparisons apply the wrong affinity & \href{https://github.com/tursodatabase/turso/issues/10342}{\#10342} & fixed \\
{}JSON5 // comment ends only at LF, not at CR, U+2028 or U+2029 & \href{https://github.com/tursodatabase/turso/issues/10343}{\#10343} & fixed \\
{}group\_concat over a sliding window frame panics or drops an empty string & \href{https://github.com/tursodatabase/turso/issues/10352}{\#10352} & reported \\
{}Column qualified with sqlite\_master not found when FROM names sqlite\_schema & \href{https://github.com/tursodatabase/turso/issues/10353}{\#10353} & reported \\
{}PRAGMA index\_xinfo reports the rowid entry's name as '' instead of NULL & \href{https://github.com/tursodatabase/turso/issues/10354}{\#10354} & reported \\
{}ANALYZE on a database without tables does not create sqlite\_stat1 & \href{https://github.com/tursodatabase/turso/issues/10355}{\#10355} & reported \\
{}soundex() returns '?000' for input with a non-letter & \href{https://github.com/tursodatabase/turso/issues/10356}{\#10356} & reported \\
{}json\_patch/jsonb\_patch decode escapes in added keys; jsonb\_patch writes a malformed key & \href{https://github.com/tursodatabase/turso/issues/10370}{\#10370} & reported \\
{}json\_error\_position one character off for a bad escape or a number without digits & \href{https://github.com/tursodatabase/turso/issues/10378}{\#10378} & reported \\
{}RENAME COLUMN of a WITHOUT ROWID primary-key column moves it to the end of the key & \href{https://github.com/tursodatabase/turso/issues/10387}{\#10387} & reported \\
{}Correlated subquery reads NULL for a hash-build table when the probe side is a LEFT JOIN & \href{https://github.com/tursodatabase/turso/issues/10388}{\#10388} & reported \\
{}subquery not found for a WHERE subquery reading a table outside a materialized hash build & \href{https://github.com/tursodatabase/turso/issues/10389}{\#10389} & reported \\
{}RIGHT JOIN unmatched-rows pass panics (cursor is None) with a nested FROM subquery & \href{https://github.com/tursodatabase/turso/issues/10390}{\#10390} & reported \\
{}ADD COLUMN ... DEFAULT -9223372036854775808 reads as REAL for existing rows & \href{https://github.com/tursodatabase/turso/issues/10391}{\#10391} & reported \\
{}STRICT type check runs before NOT NULL, so INSERT OR IGNORE fails on a row SQLite skips & \href{https://github.com/tursodatabase/turso/issues/10392}{\#10392} & reported \\
{}UPSERT DO UPDATE is compiled when the INSERT cannot conflict; its errors fail the INSERT & \href{https://github.com/tursodatabase/turso/issues/10393}{\#10393} & reported \\
{}concat\_ws() drops BLOB arguments and returns NULL when the separator is a BLOB & \href{https://github.com/tursodatabase/turso/issues/10394}{\#10394} & reported \\
{}Compound ORDER BY x ISNULL does not match a result column written x IS NULL & \href{https://github.com/tursodatabase/turso/issues/10395}{\#10395} & reported \\
{}schema.alias.column naming an outer query's table is not found in a subquery & \href{https://github.com/tursodatabase/turso/issues/10396}{\#10396} & fixed \\
{}PRAGMA table\_info/table\_list keep a view's old columns after its table changes & \href{https://github.com/tursodatabase/turso/issues/10397}{\#10397} & reported \\
{}sqlite\_schema.sql holds a re-serialized statement instead of the original text & \href{https://github.com/tursodatabase/turso/issues/10398}{\#10398} & reported \\
{}PRAGMA index\_list/index\_info/index\_xinfo omit a WITHOUT ROWID table's primary-key index & \href{https://github.com/tursodatabase/turso/issues/10399}{\#10399} & fixed \\
{}unhex(X, Y) returns NULL when Y is an INTEGER, REAL or BLOB & \href{https://github.com/tursodatabase/turso/issues/10400}{\#10400} & fixed \\
{}UPDATE of a child table fails with foreign key mismatch when no FK column is assigned & \href{https://github.com/tursodatabase/turso/issues/10401}{\#10401} & fixed \\
{}replace(X, '', NULL) returns NULL instead of X & \href{https://github.com/tursodatabase/turso/issues/10426}{\#10426} & fixed \\
{}OR over an index and the rowid misses rows: a seek leaves an earlier DeferredSeek pending & -- & not yet reported \\
{}JSON functions take a non-JSONB BLOB for JSONB and skip leading whitespace bytes & -- & not yet reported \\
{}UPDATE on a parent table fails with foreign key mismatch while foreign keys are off & -- & not yet reported \\
\addlinespace
\multicolumn{3}{@{}l}{\textbf{PostgreSQL} (open loop, 16)} \\*
{}Nested BETWEEN grows exponentially in parse analysis and ignores statement\_timeout & \href{https://www.postgresql.org/message-id/19758-d074798b35cc9215\%40postgresql.org}{BUG \#19758} & reported \\
{}Hashed set operations and recursive UNION pre-size hash tables without the hash\_mem cap & \href{https://www.postgresql.org/message-id/19759-acd3f33402e54f36\%40postgresql.org}{BUG \#19759} & reported \\
{}CLUSTER/REPACK drops a live row updated in a rolled-back subtransaction & \href{https://www.postgresql.org/message-id/19760-8f9a0eba17bc8c92@postgresql.org}{BUG \#19760} & reported \\
{}btree\_gist date index: inserting infinity fails with cannot subtract infinite dates & \href{https://www.postgresql.org/message-id/19761-b72cfc453e487610\%40postgresql.org}{BUG \#19761} & reported \\
{}EXPLAIN ANALYZE of a Parallel Append sharing a SubPlan fails an assert in shm\_toc\_insert & \href{https://www.postgresql.org/message-id/19766-c3d105a4e0ec5487@postgresql.org}{BUG \#19766} & reported \\
{}ORDER BY/GROUP BY of a negated constant deparses as a column position; dumps do not restore & \href{https://www.postgresql.org/message-id/19767-f1189fd6a83254b7\%40postgresql.org}{BUG \#19767} & reported \\
{}GROUP BY with ROLLUP and ndistinct statistics fails with corrupt MVNDistinct entry & \href{https://www.postgresql.org/message-id/19772-d8c886b4db230132\%40postgresql.org}{BUG \#19772} & reported \\
{}Cross-partition UPDATE of a self-referencing table with ON UPDATE CASCADE fails with 23503 & \href{https://www.postgresql.org/message-id/19781-a8a20b6a904d9050\%40postgresql.org}{BUG \#19781} & reported \\
{}SET CONSTRAINTS in a released savepoint survives ROLLBACK TO an outer savepoint & \href{https://www.postgresql.org/message-id/19782-02b66fb19c9ce85b\%40postgresql.org}{BUG \#19782} & reported \\
{}Sorted GROUPING SETS leave a grouping column un-nulled (PG 19 regression) & \href{https://www.postgresql.org/message-id/19783-070e4bf0c73c1076\%40postgresql.org}{BUG \#19783} & reported \\
{}min/max optimization costing = ANY over max(col) fails with no relation entry for relid 1 & \href{https://www.postgresql.org/message-id/19784-93f76236decd0656\%40postgresql.org}{BUG \#19784} & reported \\
{}JSON\_TABLE: subquery in PASSING fails with XX000; PASSING evaluated once per column & \href{https://www.postgresql.org/message-id/19786-70758103a4b2bff2\%40postgresql.org}{BUG \#19786} & reported \\
{}could not find pathkey item to sort: Gather Merge over merge join below reordered LEFT JOINs & \href{https://www.postgresql.org/message-id/19787-d14d7f0f320f4495\%40postgresql.org}{BUG \#19787} & reported \\
{}MCV statistics with a domain-over-bool column: expression not found in statistics object & BUG \#19789 & reported \\
{}XX000 cannot start a subtransaction with unprocessed inval messages after ALTER DOMAIN & BUG \#19790 & reported \\
{}ALTER COLUMN TYPE of an INCLUDE column of an index backing a foreign key fails (2BP01) & BUG \#19791 & reported \\
\addlinespace
\multicolumn{3}{@{}l}{\textbf{ada-url} (open loop, 12)} \\*
{}Drive-letter reference against a file: base marks the URL as having an opaque path & \href{https://github.com/ada-url/ada/issues/1278}{ada\#1278} & reported \\
{}url\_search\_params keeps ill-formed UTF-8 instead of decoding it to U+FFFD & \href{https://github.com/ada-url/ada/issues/1279}{ada\#1279} & declined \\
{}URLPattern test/exec/match resolve a relative string input against an empty base URL & \href{https://github.com/ada-url/ada/issues/1280}{ada\#1280} & fixed \\
{}URLPattern special-scheme pathname does not canonicalize a backslash to / & \href{https://github.com/ada-url/ada/issues/1281}{ada\#1281} & fixed \\
{}set\_host(host:port) over the max input length applies the host, drops the port, returns true & \href{https://github.com/ada-url/ada/issues/1282}{ada\#1282} & reported \\
{}canonicalize\_protocol indexes an empty string\_view for an escaped-colon protocol pattern & \href{https://github.com/ada-url/ada/issues/1283}{ada\#1283} & reported \\
{}URLPattern opaque pathname: a leading // is parsed as an authority and dropped & \href{https://github.com/ada-url/ada/issues/1284}{ada\#1284} & reported \\
{}URLPattern search canonicalization uses the non-special query encode set & \href{https://github.com/ada-url/ada/issues/1285}{ada\#1285} & reported \\
{}url\_aggregator host setters exceed the max input length with IDNA-expanding hosts & \href{https://github.com/ada-url/ada/issues/1326}{ada\#1326} & reported \\
{}generate\_pattern\_string drops braces around a part whose prefix starts with / but is longer & \href{https://github.com/ada-url/ada/issues/1327}{ada\#1327} & fixed \\
{}IDNA combining-mark, bidi and normalization tables predate the Unicode 17 mapping table & \href{https://github.com/ada-url/idna/issues/97}{idna\#97} & fixed \\
{}x86-64 builds run SSSE3 instructions without a runtime CPU check & -- & fixed \\
\addlinespace
\multicolumn{3}{@{}l}{\textbf{Cedar} (open loop, 2)} \\*
{}upsert\_entities keeps a stale ancestor when the batch names the same entity twice & \href{https://github.com/cedar-policy/cedar/issues/2606}{cedar\#2606} & fixed \\
{}add\_entities compares parents with the stored closure: drops parents, refuses identicals & \href{https://github.com/cedar-policy/cedar/issues/2629}{cedar\#2629} & reported \\
\addlinespace
\multicolumn{3}{@{}l}{\textbf{QED} (open loop, 2)} \\*
{}Real constants parsed as f32, so distinct literals are proved equal (wide ones panic) & \href{https://github.com/qed-solver/prover/issues/13}{qed\#13} & reported \\
{}Panic on INTEGER vs DECIMAL operands in IN (subquery) and UNION & \href{https://github.com/qed-solver/prover/issues/14}{qed\#14} & reported \\
\addlinespace
\multicolumn{3}{@{}l}{\textbf{uutils coreutils} (closed loop, 45)} \\*
{}comm compares lines including their newline, rejecting sorted input & \href{https://github.com/uutils/coreutils/issues/15439}{\#15439} & fixed \\
{}comm -{}-check-order -{}-total still prints the total after an order failure & \href{https://github.com/uutils/coreutils/issues/15254}{\#15254} & duplicate \\
{}cksum with a legacy algorithm omits the file name for an explicit - & \href{https://github.com/uutils/coreutils/issues/15254}{\#15254} & duplicate \\
{}csplit -{}-suppress-matched rejects a line number one past the end of input & \href{https://github.com/uutils/coreutils/issues/15584}{\#15584} & fixed \\
{}csplit /regex/ after a line-number pattern never tests the line the split stopped at & \href{https://github.com/uutils/coreutils/issues/15440}{\#15440} & reported \\
{}expand rejects -{}-tabs=2,/2 and -{}-tabs=4,+2 as not ascending & \href{https://github.com/uutils/coreutils/issues/15579}{\#15579} & fixed \\
{}dd status=none still warns that it cannot skip on a short stdin & \href{https://github.com/uutils/coreutils/pull/13115}{PR \#13115} & duplicate \\
{}fmt puts one space instead of two after a sentence-ending first word at line end & \href{https://github.com/uutils/coreutils/issues/15585}{\#15585} & reported \\
{}expand misplaces tab stops after a file without a trailing newline and past 8 KiB & \href{https://github.com/uutils/coreutils/issues/15656}{\#15656} & reported \\
{}fold counts control characters such as ESC and DEL as zero columns & \href{https://github.com/uutils/coreutils/issues/15571}{\#15571} & reported \\
{}join orders a missing join field before an empty one & \href{https://github.com/uutils/coreutils/issues/15254}{\#15254} & won't report \\
{}join treats an empty or blank-only line as one empty field & \href{https://github.com/uutils/coreutils/issues/15254}{\#15254} & won't report \\
{}nl pads unnumbered lines with width+1 spaces, ignoring the -s length & \href{https://github.com/uutils/coreutils/issues/15591}{\#15591} & reported \\
{}od -S with -Ax prints 7-digit hex offsets instead of 6 & \href{https://github.com/uutils/coreutils/issues/15592}{\#15592} & fixed \\
{}sort -V ignores -i, -d and -f & \href{https://github.com/uutils/coreutils/issues/15587}{\#15587} & reported \\
{}seq -w pads to the width of the increment, which is never printed & \href{https://github.com/uutils/coreutils/issues/15572}{\#15572} & fixed \\
{}sort ignores a key's own r when the key is the whole line (-k1r) & \href{https://github.com/uutils/coreutils/issues/15570}{\#15570} & fixed \\
{}sort treats form feed and carriage return as blanks & \href{https://github.com/uutils/coreutils/issues/15588}{\#15588} & fixed \\
{}numfmt -{}-invalid: one invalid field leaves the whole line unconverted & \href{https://github.com/uutils/coreutils/issues/15575}{\#15575} & reported \\
{}split -{}-verbose prints nothing with -n/-{}-number & \href{https://github.com/uutils/coreutils/issues/15583}{\#15583} & reported \\
{}split accepts a suffix start too large for the suffix length on empty input & \href{https://github.com/uutils/coreutils/issues/15569}{\#15569} & reported \\
{}join: trailing blanks start an extra field; -i prints the first line's join field & \href{https://github.com/uutils/coreutils/issues/15576}{\#15576} & fixed \\
{}split -a 0 with -n N uses a 1-character suffix instead of the default & \href{https://github.com/uutils/coreutils/issues/15581}{\#15581} & fixed \\
{}tr validates string2 constructs against the wrong mode & \href{https://github.com/uutils/coreutils/issues/15589}{\#15589} & reported \\
{}numfmt -{}-to=si/iec ignores -{}-round for values that get no suffix & \href{https://github.com/uutils/coreutils/issues/15574}{\#15574} & reported \\
{}sort -h ignores -f, so lowercase suffixes are not folded & \href{https://github.com/uutils/coreutils/issues/15577}{\#15577} & reported \\
{}expand counts a NUL byte as one column, so the next tab stop comes early & \href{https://github.com/uutils/coreutils/issues/15580}{\#15580} & reported \\
{}nl resets the header/footer section to body at each input file & \href{https://github.com/uutils/coreutils/issues/15586}{\#15586} & fixed \\
{}numfmt -{}-round changes the last digit of a number that needs no rounding & -- & won't report \\
{}sort: global -i is not overridden by -d & \href{https://github.com/uutils/coreutils/issues/15578}{\#15578} & reported \\
{}basenc -{}-base64url -d rejects input whose last group has no = padding\textsuperscript{*} & \href{https://github.com/uutils/coreutils/issues/15664}{\#15664} & reported \\
{}cat -s drops a single blank line after a line continued from an earlier file or read\textsuperscript{*} & \href{https://github.com/uutils/coreutils/issues/15655}{\#15655} & reported \\
{}chmod rejects modes such as -=r, -+x and -,u+x as unknown options\textsuperscript{*} & \href{https://github.com/uutils/coreutils/issues/15665}{\#15665} & reported \\
{}chmod -{}-reference processes the first FILE operand last\textsuperscript{*} & \href{https://github.com/uutils/coreutils/issues/15666}{\#15666} & reported \\
{}csplit hangs on /REGEX/-N after a line-number pattern that stopped at a regex line\textsuperscript{*} & \href{https://github.com/uutils/coreutils/issues/15653}{\#15653} & reported \\
{}cut -{}-output-delimiter prints the last field twice when the line has no terminator\textsuperscript{*} & \href{https://github.com/uutils/coreutils/issues/15654}{\#15654} & reported \\
{}dd conv=swab pairs bytes within each input block, so output depends on bs= and reads\textsuperscript{*} & \href{https://github.com/uutils/coreutils/issues/15646}{\#15646} & reported \\
{}numfmt prints -0 for a negative number that rounds to zero\textsuperscript{*} & \href{https://github.com/uutils/coreutils/issues/15661}{\#15661} & reported \\
{}printf \%e/\%g round an exact halfway value up instead of to even\textsuperscript{*} & \href{https://github.com/uutils/coreutils/issues/15662}{\#15662} & reported \\
{}printf \%.*s accepts a * precision above 2147483647 instead of failing\textsuperscript{*} & \href{https://github.com/uutils/coreutils/issues/15663}{\#15663} & reported \\
{}split -l parses its count like a byte SIZE (-l 010 means 8 lines, -l 1k accepted)\textsuperscript{*} & \href{https://github.com/uutils/coreutils/issues/15660}{\#15660} & reported \\
{}expand/unexpand take a bare - operand as an obsolete -N tab option, skipping stdin\textsuperscript{*} & \href{https://github.com/uutils/coreutils/issues/15658}{\#15658} & reported \\
{}unexpand handles a line continuing from one FILE into the next as two lines\textsuperscript{*} & \href{https://github.com/uutils/coreutils/issues/15657}{\#15657} & reported \\
{}uniq -f treats carriage return and form feed as field separators\textsuperscript{*} & \href{https://github.com/uutils/coreutils/issues/15659}{\#15659} & reported \\
{}unexpand counts a NUL byte as one column, so blanks after it hit the wrong tab stops\textsuperscript{*} & \href{https://github.com/uutils/coreutils/pull/15627}{PR \#15627} & duplicate \\
\addlinespace
\multicolumn{3}{@{}l}{\textbf{brush} (closed loop, 27)} \\*
{}+= on a declare -i variable panics on i64 overflow (overflow-checked builds) & \href{https://github.com/reubeno/brush/issues/1577}{\#1577} & reported \\
{}\$\{s/pat/rep\}: \$1, \$0, \$\{name\} in the replacement are read as regex group references & -- & fixed \\
{}Negative subscript on a sparse array counts from the element count, not the highest index & \href{https://github.com/reubeno/brush/issues/1570}{\#1570} & fixed \\
{}printf invalid-number messages ignore 2> redirection, and the exit status is 0 & \href{https://github.com/reubeno/brush/issues/1589}{\#1589} & reported \\
{}break/continue do not track how many loops enclose them & \href{https://github.com/reubeno/brush/issues/1595}{\#1595} & reported \\
{}Unexpanded \textasciitilde{}01 loses its leading zero and prints as \textasciitilde{}1 & \href{https://github.com/reubeno/brush/issues/1592}{\#1592} & reported \\
{}Error inside a compound command with 2> ignores that redirect & \href{https://github.com/reubeno/brush/issues/1596}{\#1596} & reported \\
{}Command substitution hangs when a builtin writes over 64 KiB after a nested \$( ) & \href{https://github.com/reubeno/brush/issues/1576}{\#1576} & reported \\
{}Glob "".* (empty quotes before the leading dot) does not match dot files & \href{https://github.com/reubeno/brush/issues/1584}{\#1584} & reported \\
{}An error on one line of a multi-line \$( ) does not end the command substitution & \href{https://github.com/reubeno/brush/issues/1593}{\#1593} & reported \\
{}\$\{v:=word\} expands to word, not v's new value, under declare -u/-l/-c/-i & \href{https://github.com/reubeno/brush/issues/1585}{\#1585} & reported \\
{}Brace expansion is not applied to assignment arguments of declare/local/export & \href{https://github.com/reubeno/brush/issues/1569}{\#1569} & reported \\
{}A shell error in one pipeline stage aborts the parent shell & \href{https://github.com/reubeno/brush/pull/1538}{PR \#1538} & duplicate \\
{}\$\{a[@]+x\} with an empty array gives one empty word instead of none & \href{https://github.com/reubeno/brush/issues/1568}{\#1568} & fixed \\
{}\textasciitilde{}- expands to an empty string when OLDPWD is declared but unset & \href{https://github.com/reubeno/brush/issues/1591}{\#1591} & reported \\
{}An assignment error inside a function does not abandon the rest of the calling line & \href{https://github.com/reubeno/brush/issues/1594}{\#1594} & reported \\
{}nullglob removes a non-pattern word like [a/b] (bracket expression containing /) & \href{https://github.com/reubeno/brush/issues/1582}{\#1582} & reported \\
{}echo -n -n hi fails with a usage error, and -e wins over a later -E\textsuperscript{*} & \href{https://github.com/reubeno/brush/issues/1588}{\#1588} & reported \\
{}[[ -z == x ]] is accepted as a string comparison; bash rejects it\textsuperscript{*} & \href{https://github.com/reubeno/brush/issues/1587}{\#1587} & reported \\
{}shopt -s globstar is accepted, but ** does not match recursively\textsuperscript{*} & \href{https://github.com/reubeno/brush/issues/1581}{\#1581} & reported \\
{}printf with a large * width aborts and drops the rest of the output\textsuperscript{*} & \href{https://github.com/reubeno/brush/issues/1590}{\#1590} & reported \\
{}read drops control characters from a pipe or file and treats \textasciicircum{}C/\textasciicircum{}D as keystrokes\textsuperscript{*} & \href{https://github.com/reubeno/brush/issues/1579}{\#1579} & reported \\
{}declare -f indents lines inside multi-line strings and here-documents\textsuperscript{*} & \href{https://github.com/reubeno/brush/issues/1580}{\#1580} & reported \\
{}Glob path component without metacharacters keeps backslashes from an expansion\textsuperscript{*} & \href{https://github.com/reubeno/brush/issues/1583}{\#1583} & reported \\
{}\$\{v/pat/rep\}: unquoted \& in the replacement is not replaced by the match\textsuperscript{*} & \href{https://github.com/reubeno/brush/issues/1586}{\#1586} & reported \\
{}A syntax error later in a script or -c string stops every earlier command from running\textsuperscript{*} & \href{https://github.com/reubeno/brush/issues/1578}{\#1578} & reported \\
{}( case a in a) echo y;; esac ) is a syntax error: esac ) parsed as another case item\textsuperscript{*} & \href{https://github.com/reubeno/brush/pull/1529}{PR \#1529} & duplicate \\
\addlinespace
\multicolumn{3}{@{}l}{\textbf{gitoxide} (closed loop, 31)} \\*
{}@\{push\} fails when the branch has no configured remote; git falls back to origin & \href{https://github.com/GitoxideLabs/gitoxide/issues/3096}{\#3096} & reported \\
{}@\{u\}@\{1\} and @\{push\}@\{...\} are rejected with unconsumed input & \href{https://github.com/GitoxideLabs/gitoxide/issues/3098}{\#3098} & reported \\
{}rev-parse of a short name fails when an earlier DWIM candidate is a dangling symref & \href{https://github.com/GitoxideLabs/gitoxide/issues/3090}{\#3090} & reported \\
{}<name>@\{n\} fails when the branch name looks like a short hash & \href{https://github.com/GitoxideLabs/gitoxide/issues/3099}{\#3099} & reported \\
{}@\{-N\} peels an annotated tag that git returns unpeeled & \href{https://github.com/GitoxideLabs/gitoxide/issues/3095}{\#3095} & reported \\
{}<ref>@\{n\} on an unborn HEAD or dangling symref returns a reflog entry; git fails & \href{https://github.com/GitoxideLabs/gitoxide/issues/3089}{\#3089} & reported \\
{}@\{-1\} resolves to a deleted branch whose name matches a dangling symref & \href{https://github.com/GitoxideLabs/gitoxide/issues/3090}{\#3090} & reported \\
{}SHA-256 repositories: loose objects are never found by an abbreviated id & \href{https://github.com/GitoxideLabs/gitoxide/issues/3101}{\#3101} & reported \\
{}<rev>\textasciicircum{}\{/regex\} on an unpeelable ref (unborn HEAD) searches all refs instead of failing & \href{https://github.com/GitoxideLabs/gitoxide/issues/3100}{\#3100} & reported \\
{}<ref>@\{n\} resolves to the null object id when the entry records a ref deletion & \href{https://github.com/GitoxideLabs/gitoxide/issues/3089}{\#3089} & reported \\
{}A short object prefix fails when a dangling symref has the same name & \href{https://github.com/GitoxideLabs/gitoxide/issues/3090}{\#3090} & reported \\
{}<ref>@\{0\} returns the newest reflog entry instead of the ref's current value & \href{https://github.com/GitoxideLabs/gitoxide/issues/3089}{\#3089} & reported \\
{}<ref>@\{N\} fails when the reflog has exactly N entries (oldest has a previous value) & \href{https://github.com/GitoxideLabs/gitoxide/issues/3089}{\#3089} & reported \\
{}@\{push\} with push.default=upstream and a separate push remote maps through the wrong remote & \href{https://github.com/GitoxideLabs/gitoxide/issues/3096}{\#3096} & reported \\
{}Tree merge debug-asserts: one side renames and edits a file, the other makes the same edit\textsuperscript{*} & \href{https://github.com/GitoxideLabs/gitoxide/issues/3091}{\#3091} & reported \\
{}Commit::describe() ignores annotated tags that point to another tag\textsuperscript{*} & \href{https://github.com/GitoxideLabs/gitoxide/issues/3103}{\#3103} & reported \\
{}topo traversal returns a tip that an end hides (Parents::First, or tip also an end)\textsuperscript{*} & \href{https://github.com/GitoxideLabs/gitoxide/issues/3105}{\#3105} & reported \\
{}rev\_walk().with\_hidden() returns ancestors of hidden commits under skewed commit times\textsuperscript{*} & \href{https://github.com/GitoxideLabs/gitoxide/issues/3104}{\#3104} & reported \\
{}<rev>:file/ resolves to the blob; git rejects a trailing slash on a non-directory\textsuperscript{*} & \href{https://github.com/GitoxideLabs/gitoxide/issues/3102}{\#3102} & reported \\
{}Tree merge keeps the ancestor when one side edits a file, the other makes it a symlink\textsuperscript{*} & \href{https://github.com/GitoxideLabs/gitoxide/issues/3107}{\#3107} & reported \\
{}Tree merge: same content on both sides plus one-sided chmod +x keeps the ancestor\textsuperscript{*} & \href{https://github.com/GitoxideLabs/gitoxide/issues/3097}{\#3097} & reported \\
{}status() fails when HEAD's branch is a symbolic ref to an unborn branch\textsuperscript{*} & \href{https://github.com/GitoxideLabs/gitoxide/issues/3112}{\#3112} & reported \\
{}status() misses a staged deletion when the index entry is intent-to-add\textsuperscript{*} & \href{https://github.com/GitoxideLabs/gitoxide/issues/3115}{\#3115} & reported \\
{}status() lists a file that replaced a submodule as untracked as well as a type change\textsuperscript{*} & \href{https://github.com/GitoxideLabs/gitoxide/issues/3113}{\#3113} & reported \\
{}Tree merge reports a false dir/file conflict when the other side deletes part of the dir\textsuperscript{*} & \href{https://github.com/GitoxideLabs/gitoxide/issues/3106}{\#3106} & reported \\
{}status ignores a gitlink whose path is not listed in .gitmodules\textsuperscript{*} & \href{https://github.com/GitoxideLabs/gitoxide/issues/3114}{\#3114} & reported \\
{}status() reports an executable replaced by a symlink as Modified, not TypeChange\textsuperscript{*} & \href{https://github.com/GitoxideLabs/gitoxide/issues/3111}{\#3111} & reported \\
{}Tree merge loses a file renamed onto a directory that the other side modified\textsuperscript{*} & \href{https://github.com/GitoxideLabs/gitoxide/issues/3109}{\#3109} & reported \\
{}Tree merge treats d as renamed to y after a move of d/e to y/z\textsuperscript{*} & \href{https://github.com/GitoxideLabs/gitoxide/issues/3108}{\#3108} & reported \\
{}Text merge reports no conflict when both sides append different lines at the end\textsuperscript{*} & \href{https://github.com/GitoxideLabs/gitoxide/issues/3110}{\#3110} & reported \\
{}Tree merge drops a file added on both sides when only one side makes it executable\textsuperscript{*} & \href{https://github.com/GitoxideLabs/gitoxide/issues/3097}{\#3097} & reported \\
\addlinespace
\multicolumn{3}{@{}l}{\textbf{DuckDB} (closed loop, 29)} \\*
{}Outer join with ON-clause subqueries under an inequality join: DataChunk::Slice assertion & \href{https://github.com/duckdb/duckdb/issues/27717}{\#27717} & reported \\
{}SEMI join ON true replaced by a cross product leaves a stale projection map (bind error) & \href{https://github.com/duckdb/duckdb/issues/27452}{\#27452} & reported \\
{}Nested loop join ignores the projection map after an outer join with an ON subquery & \href{https://github.com/duckdb/duckdb/issues/27717}{\#27717} & reported \\
{}Piecewise merge join ignores the projection map after an outer join with an ON subquery & \href{https://github.com/duckdb/duckdb/issues/27717}{\#27717} & reported \\
{}dense\_rank() over an empty local sort collection segfaults & \href{https://github.com/duckdb/duckdb/issues/27448}{\#27448} & fixed \\
{}Hash join with a zero-column probe side fails: tuple width should not be 0 & \href{https://github.com/duckdb/duckdb/issues/27450}{\#27450} & fixed \\
{}Join order optimizer drops a relation-free filter (WHERE NULL returns a row) & \href{https://github.com/duckdb/duckdb/issues/27451}{\#27451} & fixed \\
{}Statistics propagation table filter loses a column reference (assertion) & \href{https://github.com/duckdb/duckdb/issues/27447}{\#27447} & reported \\
{}Join elimination asserts on a conditionless join from a correlated aggregate subquery & \href{https://github.com/duckdb/duckdb/issues/27445}{\#27445} & reported \\
{}Join elimination leaves a stale parent projection map (missing child column) & \href{https://github.com/duckdb/duckdb/issues/27449}{\#27449} & reported \\
{}Statistics propagation folds an aggregate with no expressions (assertion exprs.size() > 0) & \href{https://github.com/duckdb/duckdb/issues/27446}{\#27446} & fixed \\
{}SEMI JOIN on a right-only condition returns no rows with filter\_pushdown disabled\textsuperscript{*} & \href{https://github.com/duckdb/duckdb/issues/27633}{\#27633} & reported \\
{}Caching-operator flush continues into a finished operator: join\_hashtable assertion\textsuperscript{*} & \href{https://github.com/duckdb/duckdb/issues/27722}{\#27722} & reported \\
{}LEFT JOIN ON false AND x IN (SELECT ...) fails: Failed to bind column reference\textsuperscript{*} & \href{https://github.com/duckdb/duckdb/issues/27719}{\#27719} & reported \\
{}Join with an ungrouped aggregate over an empty table: LIMIT 1 returns 3 rows\textsuperscript{*} & \href{https://github.com/duckdb/duckdb/issues/27714}{\#27714} & reported \\
{}Table filter found to be true-or-NULL keeps rows where the predicate is NULL\textsuperscript{*} & \href{https://github.com/duckdb/duckdb/issues/27715}{\#27715} & reported \\
{}Lateral reference in a FULL JOIN ON subquery: INTERNAL !ContainsDependentJoin\textsuperscript{*} & \href{https://github.com/duckdb/duckdb/issues/27716}{\#27716} & reported \\
{}CROSS JOIN + LEFT JOIN ON (SELECT true) + WHERE: assertion "normalized" in join order\textsuperscript{*} & \href{https://github.com/duckdb/duckdb/issues/27723}{\#27723} & reported \\
{}x IN (<correlated scalar subquery>, x) treated as NULL-propagating: drops a row\textsuperscript{*} & \href{https://github.com/duckdb/duckdb/issues/27712}{\#27712} & reported \\
{}FULL JOIN with an ON subquery over a CROSS JOIN: Vector::Reference type mismatch\textsuperscript{*} & \href{https://github.com/duckdb/duckdb/issues/27726}{\#27726} & reported \\
{}abs() statistics on FLOAT/DOUBLE are backwards when the range ends at 0 (wrong result)\textsuperscript{*} & \href{https://github.com/duckdb/duckdb/issues/27711}{\#27711} & reported \\
{}Join condition statistics narrowed by a sibling condition (debug\_verify\_stats mismatch)\textsuperscript{*} & \href{https://github.com/duckdb/duckdb/issues/27725}{\#27725} & reported \\
{}Lambda inside an aggregate inside a lambda: BoundReferenceExpression INTERNAL error\textsuperscript{*} & \href{https://github.com/duckdb/duckdb/issues/27720}{\#27720} & reported \\
{}SIMILAR TO / regexp filter above a FULL JOIN fails a string\_stats assertion\textsuperscript{*} & \href{https://github.com/duckdb/duckdb/issues/27724}{\#27724} & reported \\
{}IN subquery in JOIN ON plus LEFT JOIN ON subquery: index out of range INTERNAL error\textsuperscript{*} & \href{https://github.com/duckdb/duckdb/issues/27606}{\#27606} & duplicate \\
{}Outer join max cardinality set to the product: LIMIT 1 returns 2 rows, EXISTS is false\textsuperscript{*} & \href{https://github.com/duckdb/duckdb/issues/27641}{\#27641} & duplicate \\
{}Correlated IN/NOT IN over mostly-NULL subquery values takes quadratic time\textsuperscript{*} & -- & won't report \\
{}Query with streaming delim CTEs deadlocks with all threads idle (regression)\textsuperscript{*} & \href{https://github.com/duckdb/duckdb/issues/27737}{\#27737} & reported \\
{}EXPLAIN (SQL) exports a CSE-hoisted constant as a quantile argument (binder error)\textsuperscript{*} & \href{https://github.com/duckdb/duckdb/issues/27776}{\#27776} & reported \\
\addlinespace
\multicolumn{3}{@{}l}{\textbf{DataFusion} (closed loop, 20)} \\*
{}UNION with an EXISTS arm fails: physical input schema differs (stale nullability) & \href{https://github.com/apache/datafusion/pull/26201}{PR \#26201} & duplicate \\
{}GROUP BY c, CAST(c AS BIGINT) fails in eliminate\_group\_by\_constant (duplicate names) & \href{https://github.com/apache/datafusion/pull/26356}{PR \#26356} & duplicate \\
{}ORDER BY ... LIMIT 0 panics in TopK (k > 0) when eliminate\_limit does not run & \href{https://github.com/apache/datafusion/issues/26431}{\#26431} & reported \\
{}Correlated EXISTS/IN selecting CAST of its correlated column fails (duplicate names) & \href{https://github.com/apache/datafusion/issues/26033}{\#26033} & duplicate \\
{}ORDER BY col, scalar subquery LIMIT over a RIGHT JOIN fails SanityCheckPlan (regression) & \href{https://github.com/apache/datafusion/issues/26572}{\#26572} & reported \\
{}= ANY (subquery) in a JOIN ON over an aliased table fails: No field named & \href{https://github.com/apache/datafusion/issues/26573}{\#26573} & reported \\
{}Join equivalence drops or keeps side constants wrongly; ORDER BY skipped after RIGHT JOIN & \href{https://github.com/apache/datafusion/issues/26036}{\#26036} & duplicate \\
{}Filter on a constant column is pushed below a global aggregate, returning a wrong row & \href{https://github.com/apache/datafusion/issues/26545}{\#26545} & duplicate \\
{}SanityCheckPlan with sort-merge join when a filter makes a sort or join key constant\textsuperscript{*} & \href{https://github.com/apache/datafusion/issues/26624}{\#26624} & reported \\
{}SanityCheckPlan for two window functions with target\_partitions = 1\textsuperscript{*} & \href{https://github.com/apache/datafusion/issues/26626}{\#26626} & reported \\
{}push\_down\_filter substitutes the wrong expression when col and CAST(col) share a name\textsuperscript{*} & \href{https://github.com/apache/datafusion/issues/26619}{\#26619} & reported \\
{}count(*) OVER () sees only the LIMIT rows under ORDER BY ... LIMIT over another window\textsuperscript{*} & \href{https://github.com/apache/datafusion/issues/26617}{\#26617} & reported \\
{}Panic: two window functions sharing ORDER BY a + 1 get different sort keys (debug)\textsuperscript{*} & \href{https://github.com/apache/datafusion/issues/26620}{\#26620} & reported \\
{}ORDER BY <double> LIMIT n drops -0.0 rows when 0.0 arrives earlier (TopK filter)\textsuperscript{*} & \href{https://github.com/apache/datafusion/issues/26618}{\#26618} & reported \\
{}SanityCheckPlan with sort-merge join when a filter says a + b = a\textsuperscript{*} & \href{https://github.com/apache/datafusion/issues/26623}{\#26623} & reported \\
{}SanityCheckPlan for UNION ALL ... ORDER BY when each arm selects a column twice\textsuperscript{*} & \href{https://github.com/apache/datafusion/issues/26625}{\#26625} & reported \\
{}GROUP BY CAST(a AS VARCHAR), a with a common subexpression fails in optimize\_projections\textsuperscript{*} & \href{https://github.com/apache/datafusion/issues/26621}{\#26621} & reported \\
{}x > ALL (subquery) in the SELECT list fails with a schema error unless aliased\textsuperscript{*} & \href{https://github.com/apache/datafusion/issues/26622}{\#26622} & reported \\
{}ORDER BY a * 3 > 1 LIMIT n over a join fails SanityCheckPlan (Null-typed range)\textsuperscript{*} & \href{https://github.com/apache/datafusion/issues/26430}{\#26430} & duplicate \\
{}Sort-merge join claims the right side's ordering when the left key is constant\textsuperscript{*} & \href{https://github.com/apache/datafusion/issues/25052}{\#25052} & duplicate \\
\addlinespace
\multicolumn{3}{@{}l}{\textbf{binaryen} (compiler trial, 18)} \\*
{}-{}-translate-to-exnref produces invalid IR for a delegate target ending in a none-typed instr & \href{https://github.com/WebAssembly/binaryen/issues/9291}{binaryen\#9291} & reported \\
{}Binary reader: element segment indexes off by one after a declarative segment & -- & not yet reported \\
{}Text parser leaves declarative element segments out of the elem index space & -- & not yet reported \\
{}Text parser rejects a table init expression written as plain instructions & -- & not yet reported \\
{}memory.copy between 32-bit and 64-bit memories aborts the reader & -- & not yet reported \\
{}Readers reject an if with an unreachable condition targeted by a branch inside it & -- & not yet reported \\
{}Readers reject an if with block parameters after unreachable & -- & not yet reported \\
{}Readers abort with length\_error on a multivalue if with an unreachable condition & -- & not yet reported \\
{}Readers hit an assertion in fixExtraOutput for a loop with params & -- & not yet reported \\
{}Text parser rejects a labeled if whose condition sets a non-nullable local & -- & not yet reported \\
{}OptimizeInstructions treats two identical suspends as equal (silent miscompile) & -- & not yet reported \\
{}-{}-outlining treats instructions with different operand types as equal & -- & not yet reported \\
{}-{}-outlining outlines multi-result sequences without multivalue & -- & not yet reported \\
{}-{}-outlining aborts on unreachable atomic.rmw, struct.get, call\_ref & -- & not yet reported \\
{}-{}-outlining asserts in StackSignature on repeated code with unreachable operands & -- & not yet reported \\
{}-{}-translate-to-exnref drops a typed try's type on try\textasciitilde{}delegate & -- & not yet reported \\
{}-{}-llvm-nontrapping-fptoint-lowering asserts on an unreachable trunc\_sat operand & -- & not yet reported \\
{}-{}-memory64-lowering makes an unreachable memory.grow/table.grow invalid & -- & not yet reported \\
\addlinespace
\multicolumn{3}{@{}l}{\textbf{OCaml} (compiler trial, 2)} \\*
{}Non-exhaustive integer match crashes instead of raising Match\_failure & \href{https://github.com/ocaml/ocaml/issues/15185}{ocaml\#15185} & reported \\
{}amd64 -linscan: Atomic.compare\_and\_set returns true without storing the value & \href{https://github.com/ocaml/ocaml/issues/15189}{ocaml\#15189} & reported \\
\addlinespace
\multicolumn{3}{@{}l}{\textbf{OxCaml} (compiler trial, 10)} \\*
{}-Oclassic crashes on a division by constant zero inside a try & \href{https://github.com/oxcaml/oxcaml/issues/7668}{oxcaml\#7668} & fixed \\
{}-Oclassic crashes on a float\# field read in an unreachable match arm & \href{https://github.com/oxcaml/oxcaml/issues/7683}{oxcaml\#7683} & reported \\
{}Reaper with phantom lets: an unused local in a closure breaks compilation & \href{https://github.com/oxcaml/oxcaml/issues/7684}{oxcaml\#7684} & reported \\
{}Binary emitter encodes 32-bit bswap of r8d-r15d without REX.B & \href{https://github.com/oxcaml/oxcaml/issues/7700}{oxcaml\#7700} & reported \\
{}-use-ssa: validator rejects x <= 1 against x < 2 for the same switch & \href{https://github.com/oxcaml/oxcaml/issues/7701}{oxcaml\#7701} & reported \\
{}-use-ssa: a call before an unused pure op becomes a tail call in one pipeline only & \href{https://github.com/oxcaml/oxcaml/issues/7702}{oxcaml\#7702} & reported \\
{}-flambda2-expert-inline-effects-in-cmm: duplicate binding order on a twice-used float & \href{https://github.com/oxcaml/oxcaml/issues/7704}{oxcaml\#7704} & reported \\
{}-flambda2-expert-inline-effects-in-cmm: Array.iter2 crash in emit\_expr\_raise & \href{https://github.com/oxcaml/oxcaml/issues/7705}{oxcaml\#7705} & declined \\
{}-no-cfg-eliminate-dead-trap-handlers: regalloc check fails on a nested try & -- & won't report \\
{}-no-cfg-eliminate-dead-trap-handlers: dead loop around a try kept as a cycle & -- & won't report \\
\addlinespace
\end{longtable}
\noindent\textsuperscript{*}\,Found by a later ephyra-agent run (2026-10-21 or 2026-10-22), not in the campaign of Section 4.2 (72 bugs).\par
\noindent\textsuperscript{\dag}\,Found in review, not by a fuzzing run (6 bugs).\par
}

\end{document}